\documentclass[lettersize,journal,twoside]{IEEEtran}

\usepackage{amsmath}
\usepackage{hyperref}
\usepackage{dsfont}
\usepackage{tcolorbox}
\tcbuselibrary{theorems}
\usepackage{enumitem}
\usepackage{cite}
\usepackage{amsmath,amsfonts}
\usepackage{datetime2}
\usepackage{amsthm}
\usepackage[linesnumbered, ruled, vlined]{algorithm2e}
\usepackage{multicol}
\usepackage[export]{adjustbox}
\usepackage{multirow}
\usepackage{tabularx}
\usepackage{balance}
\usepackage{rotating}
\usepackage{pifont}
\usepackage{url}
\usepackage{siunitx}
\usepackage{subfig}
\usepackage{ragged2e}
\usepackage{array,booktabs,makecell}
\usepackage{tikz}
\usetikzlibrary{arrows.meta, positioning}
\usepackage[edges]{forest}
\usepackage{adjustbox}
\usepackage{rotating}
\usepackage{soul}

\newcolumntype{P}[1]{>{\RaggedRight\arraybackslash}p{#1}}
\def\n{{\bf M}}

\ifodd 0
  
  \newcommand{\CommentAni}[1]{\textcolor[rgb]{1,0,0}{[Ani comment: #1]}}
  \newcommand{\CommentWong}[1]{\textcolor[rgb]{1,0,0}{[Wong comment: #1]}}
  \newcommand{\hlb}[1]{\textcolor{blue}{#1}}
  
\else
  \newcommand{\CommentAni}[1]{}
  \newcommand{\CommentWong}[1]{}  
  
  \newcommand{\hlb}[1]{}
  
\fi

\begin{document}

\title{Exposing Vulnerabilities in Counterfeit Prevention Systems\\Utilizing Physically Unclonable Surface Features}

\markboth{Under Review}%
{Nakra \MakeLowercase{\textit{et al.}}: Exposing Vulnerabilities in Counterfeit Prevention Systems Utilizing Physically Unclonable Surface Features}

\author{
Anirudh Nakra,~\IEEEmembership{Graduate Student Member,~IEEE,}
Nayeeb Rashid,~\IEEEmembership{Graduate Student Member,~IEEE,}
Chau-Wai~Wong,~\IEEEmembership{Senior Member,~IEEE,}
and~Min~Wu,~\IEEEmembership{Fellow,~IEEE}%
\thanks{This work was supported in part by the US National Science Foundation (award numbers ECCS-2227261 and ECCS-2227499 \textit{(Corresponding author: Min Wu})}%
\thanks{Anirudh Nakra, Nayeeb Rashid, and Min Wu are with the Department of Electrical and Computer
Engineering and the Institute for Advanced Computer Studies, University of Maryland, College Park, MD 20742 USA.}%
\thanks{Chau-Wai Wong is with the Department of Electrical and Computer Engineering, the Forensic Science Cluster, and the Secure Computing Institute, NC State University, NC 27695 USA. 
}}

\maketitle

\begin{abstract}
    Counterfeit products pose significant risks to public health and safety through infiltrating untrusted supply chains. Among numerous anti-counterfeiting techniques, leveraging inherent, unclonable microscopic irregularities of paper surfaces is an accurate and cost-effective solution. Prior work of this approach has focused on enabling ubiquitous acquisition of these physically unclonable features (PUFs). 
However, we will show that existing authentication methods relying on paper surface PUFs may be vulnerable to adversaries, resulting in a gap between technological feasibility and secure real-world deployment. 
This gap is investigated through formalizing an operational framework for paper-PUF-based authentication. 
Informed by this framework, we reveal system-level vulnerabilities across both physical and digital domains, designing physical denial-of-service and digital forgery attacks to disrupt proper authentication. 
The effectiveness of the designed attacks underscores the strong need for security countermeasures for reliable and resilient authentication based on paper PUFs. 
The proposed framework further facilitates a comprehensive, stage-by-stage security analysis, guiding the design of future counterfeit prevention systems. 
This analysis delves into potential attack strategies, offering a foundational understanding of how various system components, such as physical features and verification processes, might be exploited by adversaries.
\end{abstract}


\section{Introduction}

\begin{figure*}
    \centering
    \subfloat[\label{fig:paper}]{%
    \includegraphics[width=0.2\linewidth]{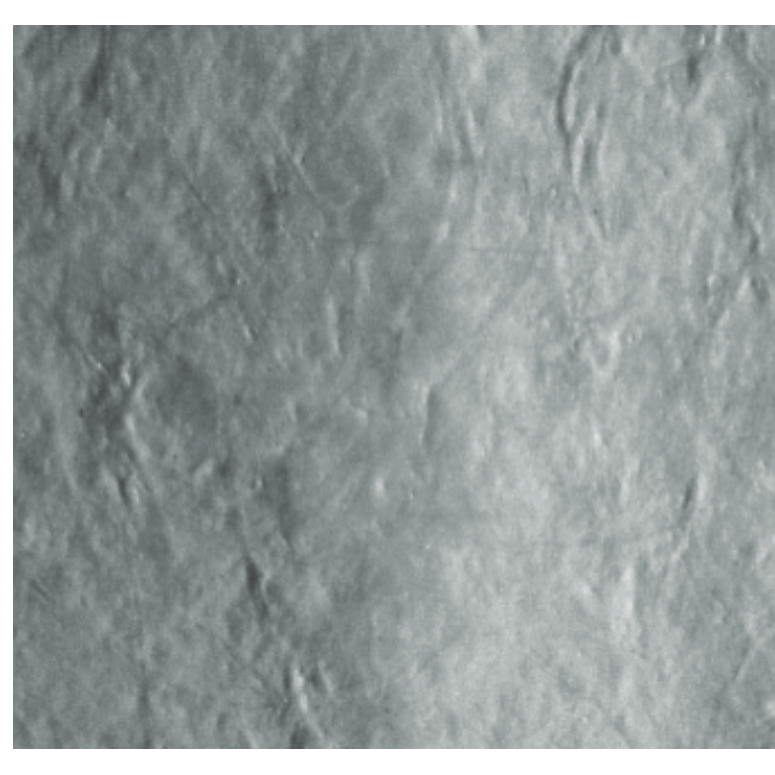}}
    \hfill
    \subfloat[\label{fig:topology}]{%
        \includegraphics[height=3.5cm,width=0.22\linewidth]{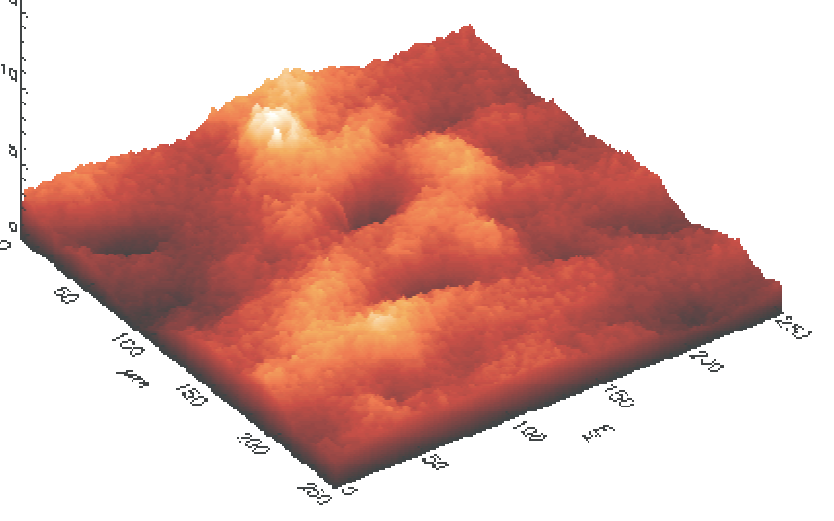}}
    \hfill
        \subfloat[\label{fig:cfm}]{%
        \includegraphics[trim={1cm 1cm 0 0},clip,width=0.2\linewidth]{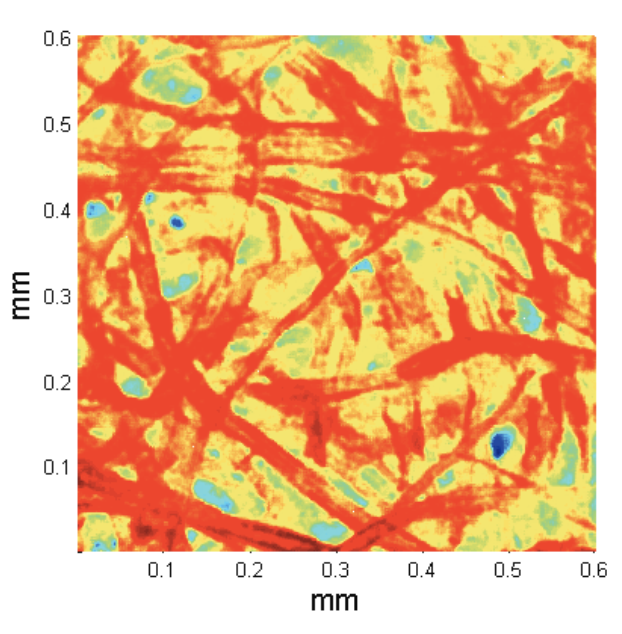}}
    \hfill
    \subfloat[\label{fig:normmap}]{%
        \includegraphics[width=0.2\linewidth]{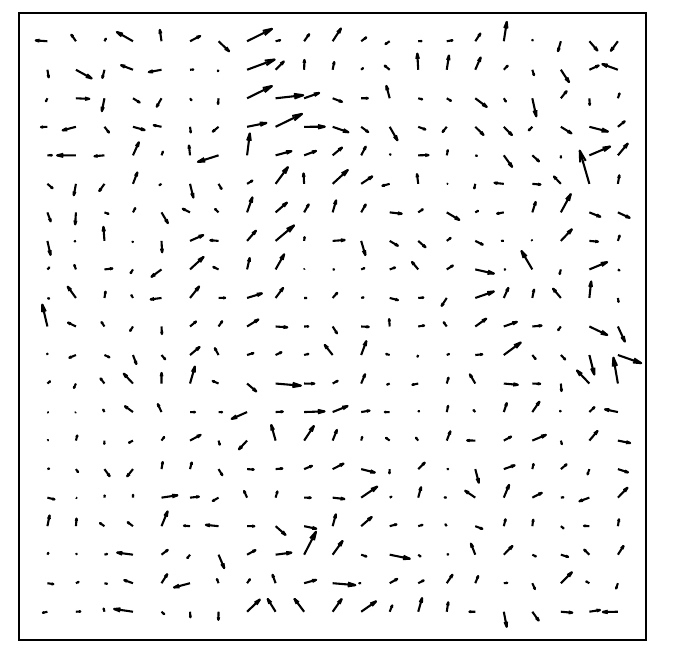}}
  \caption{\label{fig:EgPUF}Visualization of physically unclonable features (PUFs) of papers in the form of (a)~an RGB image, (b)~a topographical map, (c)~a confocal microscope scan, reproduced from \cite{rehberger2007topographical}, and (d)~a 2D projection of a matrix of 3D microscopic normal vectors (referred to as the norm map). The literature has demonstrated that the intricate microstructures of paper have strong authentication performance in counterfeit prevention systems. This work reveals critical security vulnerabilities in state-of-the-art paper-PUF-based authentication systems, which must be addressed before widespread deployment.}
  \vspace{-4mm}
\end{figure*}

Counterfeits are prevalent across critical industries, including electronics, healthcare, and automotives \cite{counterfeitcbp}. 
These counterfeit goods reach consumers through logistical distribution channels known as supply chains. Counterfeiters frequently penetrate existing untrusted supply chains
to introduce counterfeit products into the market, posing significant risks to public health and safety. Skilled forgers duplicate critical identification documents such as state-issued IDs and passports, creating substantial security threats. 
Counterfeit medications, which refer to recycled, expired, or fake medicine, were among the most seized U.S. health and safety products in 2022 \cite{counterfeitcbp}.   

Researchers have proposed various anti-counterfeiting solutions. Special inks~\cite{cox2002digital}, engravings~\cite{murataj2024artificial}, copy-resistant patterns~\cite{picard2004digital}, and optically variable features~\cite{ren2020optical} have been developed to secure products against potential forgery. 
However, these methods often require specialized verification techniques, limiting the range of stakeholders---particularly verification experts---who can participate, and introducing significant overhead to the verification process.
In contrast to the class of anti-counterfeiting solutions that add distinguishing patterns to different products, there is a growing body of literature using the intrinsic properties of materials, such as metal~\cite{FIBAR}, fabric~\cite{sharma2017fake}, and paper~\cite{toreini2017texture,beekhof2008secure,guarnera2019new,clarkson2009fingerprinting,wong2017counterfeit,flatbed}, to validate a product's authenticity. 
Paper-based authentication systems~\cite{toreini2017texture,beekhof2008secure,guarnera2019new,clarkson2009fingerprinting,wong2017counterfeit,flatbed,datta2024enabling}, 
taking advantage of microscopic imperfections in the surface texture of the paper to authenticate products, are an economical solution.
These methods may be easily deployed, as paper surfaces are commonly used in product packaging, and may be potentially applied to authenticate vital security credentials such as passports and ID cards.  

Contrary to the smooth appearance of paper surfaces to naked eyes as shown in Fig.~\ref{fig:EgPUF}(a), they are not perfectly flat. Instead, they exhibit microscopic variations, akin to the hills and valleys of mountainous regions, as visualized in Fig.~\ref{fig:EgPUF}(b). These unique structural features, formed by the intricate entanglement of wood fibers as visualized in Fig.~\ref{fig:EgPUF}(c), are intrinsic to the paper and cannot be precisely replicated by current manufacturing processes. As a result, these rich, distinct, and unclonable microscopic structures~(microstructures) can be leveraged as physically unclonable features (PUFs), serving as a unique ``fingerprint'' for each piece of paper. This inherent variability makes paper surfaces an ideal solution for preventing counterfeiting. Since paper PUFs are nearly imperceptible to naked eyes, prior research has developed algorithms to extract paper PUFs, with notable contributions from Clarkson et al.~\cite{clarkson2009fingerprinting}, Wong and Wu~\cite{wong2017counterfeit}, and Liu et al.~\cite{flatbed}, who focused on demonstrating the technological feasibility of using consumer-grade imaging devices, such as flatbed scanners and mobile cameras, for paper-PUF-based authentication. 
While significant efforts have been made to design algorithms for extracting paper PUFs, system-level security studies have been lacking---Clarkson et al. \cite{clarkson2009fingerprinting} provided preliminary threat models, and Liu et al. \cite{flatbed} explored infrastructural aspects, such as the client-server model, but neither offers a systematic security analysis for systems that adopt paper PUFs for authentication. The use of naive per-pixel correlators \cite{wong2017counterfeit,flatbed} and the absence of template protection mechanisms \cite{wong2017counterfeit,flatbed} leave state-of-the-art methods vulnerable to motivated counterfeiters. 

To establish paper PUFs as a practically deployable technique, it is important to (\textit{i})~address system-level vulnerabilities and (\textit{ii})~understand to what extent state-of-the-art paper-PUF-based authentication systems can be compromised. 
In this work, we demonstrate that paper-PUF-based counterfeit prevention systems may exhibit critical vulnerabilities at both the physical and digital levels. 
To our knowledge, no prior research has analyzed the system-level security of paper-PUF-based anti-counterfeiting systems. 
Our paper aims to fill this gap by focusing on three key efforts. 
First, we introduce an operational framework for paper-PUF-based authentication, which we build on to design novel, practically feasible attacks spanning both physical and digital domains, specifically targeting the disruption of proper authentication processes. 
Second, we provide experimental evidence demonstrating the effectiveness of these attacks, thereby exposing critical vulnerabilities that could allow counterfeit products to enter circulation. 
Third, we conduct a comprehensive, stage-by-stage analysis of our framework's vulnerabilities, capturing key system intricacies to guide future paper-PUF-based authentication system design. 
The contributions of this work are threefold.

\begin{itemize}[left=0pt]
    \item We experimentally demonstrate that \textbf{physical} denial-of-service attacks can successfully sabotage paper-PUF-based authentication systems, revealing vulnerabilities in the physical domain. 
    
    \item This study is the first to demonstrate that paper-PUF-based features can be forged in the \textbf{digital} domain, circumventing the need for the adversaries to access the authentic physical paper patches in the verification process.  

    \item We bridge the gap between the practical deployability of paper-PUF-based authentication systems and their theoretical design. We formalize an operational framework for conducting a stage-by-stage attack strategy analysis. This analysis integrates theory and best practices from signal processing, biometrics, and cryptography, offering a comprehensive and practical view of the system’s security for guiding future secure authentication system design.
\end{itemize}

The rest of this paper is organized as follows. Section~\ref{Sec2} discusses the utility of PUFs in combating counterfeiting, comparing them to traditional solutions and highlighting the advantages of paper PUFs as an economical and practical alternative. Section~\ref{Sec3} introduces an operational framework that decomposes the state-of-the-art paper-PUF-based authentication systems into their constituent components. 
Section~\ref{Sec4} exposes critical security vulnerabilities in paper-PUF-based authentication systems, by implementing a set of representative threats corresponding to two primary adversarial directions: disrupting authentication through physical denial-of-service attacks (Section~\ref{Sec4.2}) and circumventing authentication via digital forgery attacks (Section~\ref{Sec4.3}). 
Section~\ref{Sec5} leverages our operational framework to provide a holistic analysis of the system-level vulnerabilities in paper-PUF-based authentication systems. 
This serves as guidance for designers of future counterfeit prevention systems.
Finally, Section~\ref{Sec8} concludes the paper.

\vspace{-1.5mm}

\section{\label{Sec2}PUF-based-Authentication Preliminaries}

\begin{figure*}[!ht]
    \centering
         \subfloat[]{\includegraphics[width=0.55\linewidth]{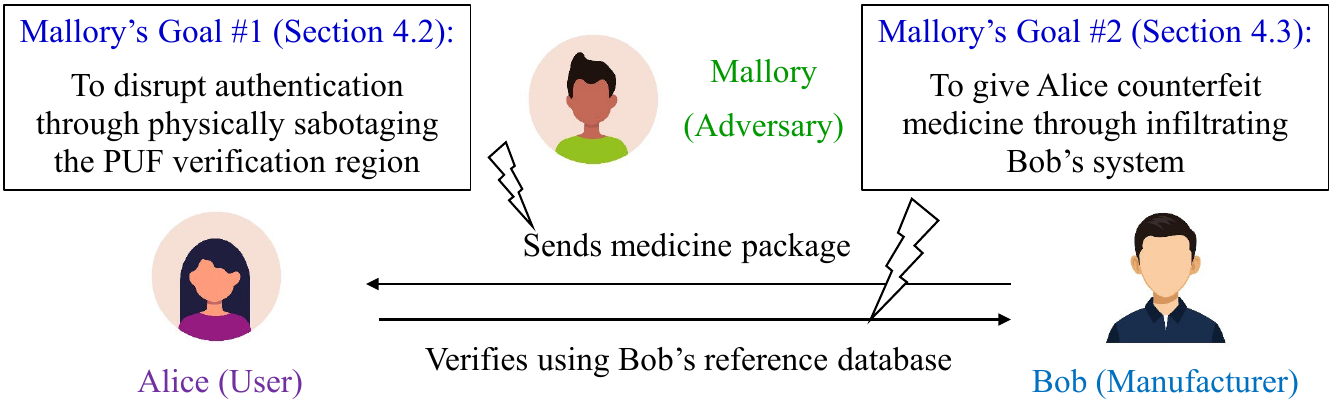}}
         \hfill \tikz{\draw[densely dashed, thick](0,3) -- (0,0);} \hfill 
        \subfloat[]{\raisebox{4ex}{\includegraphics[width=0.4\linewidth]{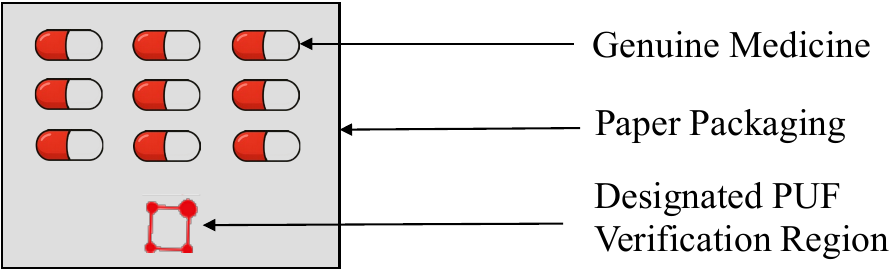}}}
    \caption{\label{fig:PharmaPUF} Illustration of a paper-PUF-based authentication system deployed in a pharmaceutical supply chain: (a)~A patient, Alice, receives a medicine package from a manufacturer, Bob. She uses an open-source app to verify the medicine's authenticity by comparing extracted PUF features with those pre-stored in Bob's reference database. Mallory, the adversary, has two primary goals: (1)~To disrupt authentication through physically sabotaging the PUF verification region, introducing ambiguity between genuine wear and tear and malicious tampering, and (2)~to authenticate counterfeit medicines as genuine through infiltrating Bob's system. (b)~A genuine medicine package, containing a PUF verification region on the paper packaging, is treated as a single inseparable entity so that the authenticity of the medicine is tied to its packaging. We use this paper-PUF-based authentication system to explore realistic adversarial possibilities.}
    \vspace{-4mm}
\end{figure*}

\subsection{Traditional Anti-Counterfeiting Solutions}

In the physical world, the tracking and tracing of a product can be performed using external identifiers such as bar codes \cite{barcode}, QR codes \cite{qrcode}, and radio frequency identification~(RFID) tags \cite{angeles2005rfid} that are attached to the product or its packaging. Even though they differ in implementation, all these identifiers can be thought of as an unencrypted container storing data that can be decoded. Due to their unencrypted designs, they are easily duplicable using consumer-grade devices such as mobile phones \cite{cloneRFID}. This implies that counterfeit products can have the same identifier as the authentic product and thus move through a supply chain undetected. To mitigate this vulnerability, copy detection patterns~(CDPs) were proposed to augment bar/QR codes to defend against duplication, but they are vulnerable to generative machine learning-based attacks \cite{slavatifs}. Some active RFID tags support built-in cryptographic protocols, but these tags are relatively expensive and bulky. Finally, these identifiers have an authentication error rate on the order of 0.01 to 0.1 \cite{counterfeitcdp1}, which is too high to be practical for a large-scale supply chain deployment. Distinct from traditional anti-counterfeiting solutions, this work focuses on counterfeit prevention systems that leverage paper PUFs, offering an economical, ubiquitous, and high-performance solution.

\subsection{Paper-PUF-based Authentication}

\subsubsection{Using Paper to Combat Counterfeiting}
Researchers have explored various special materials with unique visible features to prevent counterfeiting \cite{bubblefibertag}. 
However, many of these identifiers require proprietary infrastructure and are relatively expensive. This highlights the need for a solution that adds minimal overhead to supply chains, is easily acquired, and offers high authentication performance. 
A viable candidate that meets these criteria is optical physically unclonable functions (PUFs), which leverage the interaction of light with physical materials to generate unique features that are difficult to replicate due to the uncontrollable intrinsic properties of the material \cite{papputhesis}. While cryptographic studies have focused on using these intrinsic imperfections for random number generation \cite{yildiz2020plgakd}, researchers have also explored optical PUFs as signatures for object authentication in anti-counterfeiting systems. 
Among various optical PUFs based on materials such as metal \cite{FIBAR,ishiyama2018fast}, fabric \cite{sharma2017fake}, and paper \cite{toreini2017texture,clarkson2009fingerprinting,wong2017counterfeit}, the microscopic surface features of paper, i.e., paper PUFs, emerge as a promising solution for several reasons.

First, the interaction of light with paper surfaces can be effectively modeled using a simple diffuse reflection model, enabling straightforward PUF extraction algorithms \cite{clarkson2009fingerprinting, wongicip, wong2017counterfeit, flatbed}. 
Second, paper-PUF extraction utilizes basic acquisition hardware such as scanners \cite{clarkson2009fingerprinting, flatbed} and mobile cameras \cite{wong2017counterfeit}, minimizing overhead and simplifying deployment in supply chains. 
Third, paper-PUF-based systems achieve authentication error rates as low as $10^{-13}$ to $10^{-130}$ \cite{clarkson2009fingerprinting, wong2017counterfeit}, ensuring high performance. 
Fourth, recent research shows that paper PUF algorithms can be applied to other materials, such as IC chips \cite{liu2024surface}, despite their non-conformity to the diffuse reflection model. 
Finally, the widespread adoption of paper packaging, driven by environmental concerns, enhances the relevance and potential for real-world deployment of paper-PUF-based anti-counterfeiting systems. 
These factors collectively position paper PUFs as a viable solution for practical deployment, forming the foundation of our work.

\subsubsection{Paper PUF Extraction}
The extraction of inherent randomnesses of the paper surface has been studied by using both physical modeling and specialized imaging setups. 
Physical features characterize the surface texture of the physical material.
Clarkson et al.~\cite{clarkson2009fingerprinting} pioneer the extraction of the microscopic level texture of a paper surface as a unique signature through the use of commodity scanners to acquire four views of the image at four different orientations. 
Based on a fully diffuse reflection model from which these four images are generated, this work estimates a 2D projection [referred to as the norm map and visualized in Fig.~\ref{fig:EgPUF}(d)] of a matrix of 3D
microscopic normal vectors.
Since flatbed commodity scanners might not always be available or inconvenient, Wong and Wu~\cite{wong2017counterfeit,wongwifs} extend the estimation of the norm map to use the camera of mobile devices. 
Further work by Liu et al.~\cite{flatbed} extends Clarkson et al. to allow an additional specular component in the reflection medium.
Apart from studies that measure or estimate physical surface features, a significant body of research employs specialized imaging acquisition setups to capture high-resolution visually rendered features of the paper surface \cite{beekhof2008secure,guarnera2019new}.
However, visually rendered features have limitations as they are sensitive to capturing conditions, may require specialized imaging methods to function well, and often have weaker authentication performance than physical features~\cite{clarkson2009fingerprinting,wong2017counterfeit}. 
For this reason, this work focuses on paper-PUF-based authentication that leverages norm maps, which exemplify the state of the art due to low authentication error rates \cite{flatbed} and their use of ubiquitous sensors~\cite{wong2017counterfeit}.

\begin{figure*}[!ht]
    \centering
    \includegraphics[width=\linewidth]{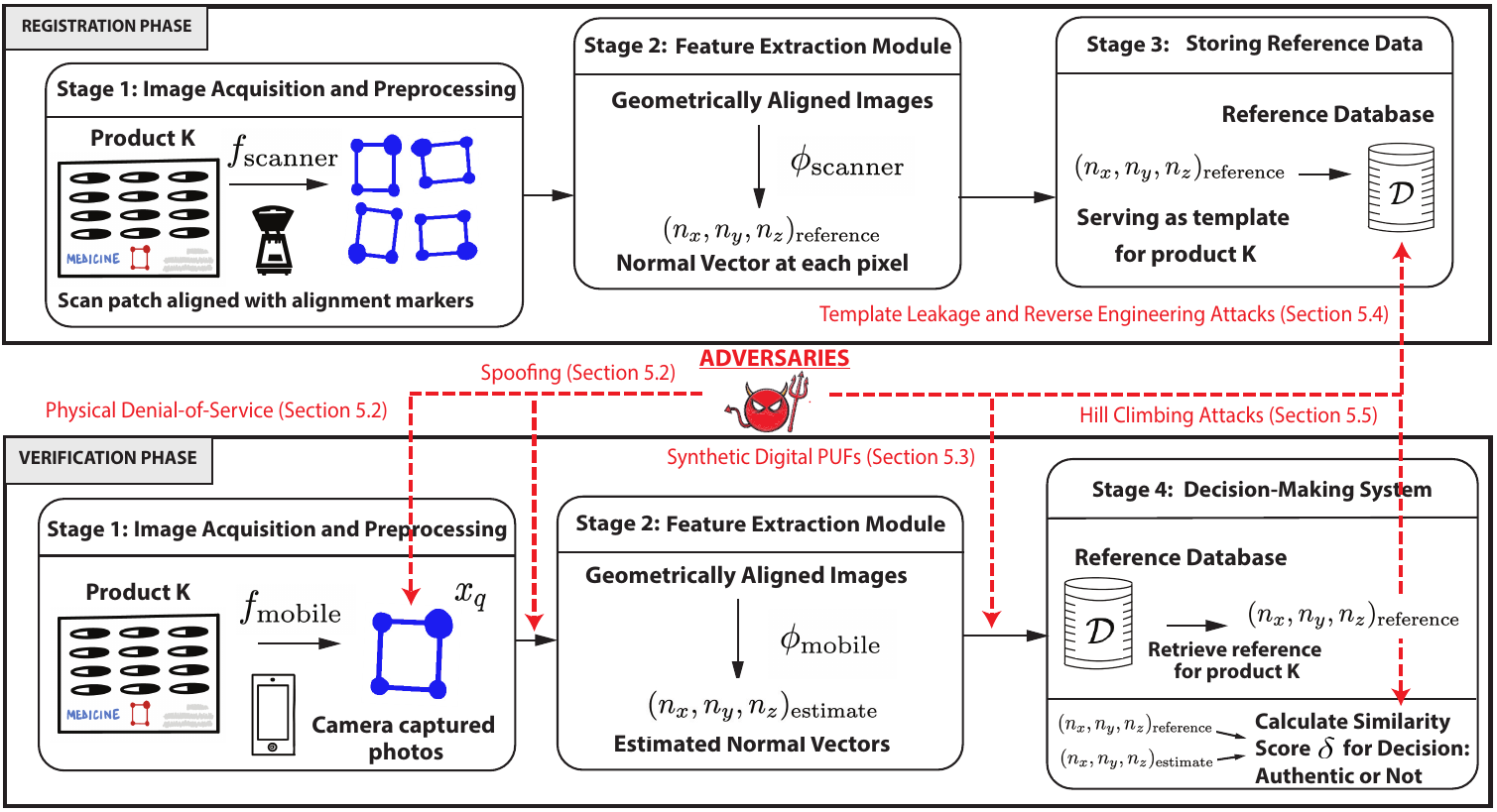}
    \caption{The norm map based anti-counterfeiting system framework $\mathcal{A} = (f,\phi,\mathcal{D},\delta) $ is divided into operational stages 1--4. We characterize the set of all potential vulnerabilities using this operational framework and highlight the most vulnerable links using red dashed arrows.
    Physical denial-of-service attacks aim to sabotage image acquisition and preprocessing. Spoofing attacks inject malicious inputs that authenticate counterfeit products as genuine. Synthetic generation attacks leverage state-of-the-art generative models to create fake features capable of passing authentication. Template leakage attacks seek to overwrite or extract stored reference features. Reverse engineering attacks estimate input images based on leaked reference features. Hill-climbing attacks iteratively adjust query features to digitally forge a PUF that successfully passes authentication.
    }
    \label{fig:framework}
    \vspace{-4mm}
\end{figure*}

\subsubsection{Paper-PUF Example in Practice: Counterfeit Medicines}

We demonstrate the deployment of a paper-PUF-based counterfeit prevention system through a pharmaceutical industry example. 
Counterfeit medications introduce harmful adulterants or ineffective chemicals that bypass regulatory certification. These adulterants can cause life-threatening conditions, while ineffective chemicals could contribute to antimicrobial resistance.
In Sub-Saharan Africa, counterfeit antimalarial and antibacterial drugs have been linked to over 500,000 deaths \cite{pharmadeath}. 
Although several PUF-based systems have recently been proposed for counterfeit medicine detection, many of these approaches depend on specialized materials \cite{murataj2024artificial,kingsley2024electrospray}, fabrication steps \cite{murataj2024artificial,kingsley2024electrospray}, or equipment, such as electrosprayed tags \cite{kingsley2024electrospray} or chemically synthesized substrates \cite{murataj2024artificial}, which can hinder scalability and increase costs in large-scale pharmaceutical supply chains.
In contrast, we demonstrate the deployment of a paper-PUF-based counterfeit prevention system through a pharmaceutical industry example, exploring a minimal infrastructural overhead solution leveraging the intrinsic microstructure of commodity paper packaging as an optical PUF, avoiding additional manufacturing complexity. 

In this context, we consider three key stakeholders in the pharmaceutical supply chain illustrated in Fig~\ref{fig:PharmaPUF}(a): Alice, Bob, and Mallory. Alice represents a legitimate patient who purchases pharmaceutical products from a reputable pharmacy, either online or physically. Bob, a pharmaceutical manufacturer, registers PUFs from the paper packaging of medicine bottles into their verification database to enable end-user authentication. Bob sends Alice a bottle of medicine wrapped in paper packaging. Upon receipt, Alice uses an official, open-source mobile app maintained by the drug company to verify the authenticity of the product by matching the extracted features to those stored in Bob's database. Mallory, the adversary, has two primary goals. 
First, she may physically disrupt the authentication process, creating ambiguity between damaged genuine and counterfeit products. Second, she may attempt to bypass the authentication process and have counterfeit products authenticated through Bob's software. Exposing vulnerabilities targeting these two goals, Section~\ref{Sec4} identifies and experimentally demonstrates the effectiveness of representative practical attacks.
To further help Bob design secure paper-PUF-based authentication systems, Section~\ref{Sec5} performs a comprehensive qualitative analysis uncovering all potential vulnerabilities.

\subsubsection{Assumptions} We define the scope of our system-level security analysis with two realistic assumptions.
First, we assume that the product and its associated packaging, as illustrated in Fig.~\ref{fig:PharmaPUF}(b), are inseparable. This assumption is supported by two main reasons: (1) PUFs are typically deployed alongside other security mechanisms, such as tamper protection seals, rather than in isolation; and (2) paper-PUF-based algorithms have shown adaptability to various surfaces, including those of IC chips~\cite{liu2024surface}. Future authentication systems may shift to capturing the microscopic features of the product itself, independent of its packaging, making our assumption forward-looking. Without this assumption, Mallory, with access to the genuine product's packaging, could remove the genuine product and replace it with a counterfeit. 

The second assumption is that the microstructures of paper surfaces are inherently stochastic and unclonable. The complex and uncontrollable intertwining of millions of cellulose fibers during paper formation \cite{sjostrom1998analytical} results in unique microstructures. This random entanglement, driven by intricate chemical and mechanical processes during the paper production, ensures that even an adversary with access to paper-making machines would be unable to replicate the microstructures of one sheet to match another. 
The difficulty of replicating the random entanglement of millions of cellulose fibers severely limits adversarial possibilities. 
In light of this, we focus on the most viable current strategy: demonstrating the feasibility of generating digital forgeries capable of deceiving the authentication application.
Future research may explore bridging this gap by translating digital forgeries into physical counterparts through high-resolution additive manufacturing or other material replication techniques.

\section{A Four-Stage Operational Framework for Paper-PUF-based Authentication Systems \label{Sec3}}

To expose vulnerabilities in counterfeit prevention systems, gaining a comprehensive, system-level understanding of state-of-the-art paper-PUF-based authentication systems is essential. 
This involves analyzing both the physical components, such as the product itself, and the digital elements, such as the feature extraction process, along with their interactions. 
Thus, we propose a general operational framework for paper-PUF-based authentication, segmenting the system into four stages, as shown in Fig.~\ref{fig:framework}, and providing the foundation for a formal adversarial analysis exposing paper PUF vulnerabilities.

\smallskip
\noindent
\textbf{Notation \& Terminology.} We denote a paper-PUF authentication system by $\mathcal{A} =(f,\phi,\mathcal{D},\delta)$, where $f$ refers to the preprocessing function that converts a query $x_{\text{q}}$ into a form appropriate for feature extraction, $\phi$ converts the preprocessed query $f(x_{\text{q}})$ into an authentication feature vector, $\mathcal{D}$ contains the reference feature vectors, and $\delta$ is the decision rule that compares $\phi(f(x_{\text{q}}))$ to the references in $\mathcal{D}$.

\smallskip
\noindent
\textbf{Registration vs. Verification Phases.}
Similar to most biometric authentication systems, a paper-PUF-based authentication system operates in two phases. The first phase is referred to as the \textit{Registration} phase, involving stages 1, 2, and 3 (top row of Fig.~\ref{fig:framework}). In this phase, the manufacturer in the supply chain extracts template norm maps from images of the surface of a product (or its paper packaging) and stores them in the reference database $\mathcal{D}$. In the second phase, known as the \textit{Verification} phase and involving stages 1, 2, and 4 (bottom row of Fig.~\ref{fig:framework}), in which the other stakeholders send queries $x_{\text{q}}$ to $\mathcal{A}$ to authenticate the various products they possess. We now explain the four operational stages in detail. 

\smallskip
\noindent
\textbf{Stage 1: Image Collection and Preprocessing Using $f$.} This stage collects and preprocesses multiple images captured from a paper patch, which will be fed into a norm map estimation algorithm at stage 2. Four or more images are acquired using scanners (left panel, top row of Fig.~\ref{fig:framework}) or cameras (left panel, bottom row of Fig.~\ref{fig:framework}) for subsequent processing. The preprocessing includes the geometric alignment of these images, which is necessary since any slight misalignment between them can cause a significant drop in the accuracy of the norm map estimation algorithms. This stage, responsible for signal acquisition and preprocessing for feature extraction, is the most vulnerable to physical denial-of-service (Sec.~\ref{Sec4.2}, Sec.~\ref{Attacks_Stage_1}) and spoofing attacks (Sec.~\ref{Attacks_Stage_1}), which aim to disrupt the process or inject malicious data into the system. 

\smallskip
\noindent
\textbf{Stage 2: Feature Extraction via $\phi$.} The feature extraction module, as illustrated in the middle panel of both rows of Fig.~\ref{fig:framework}, implements an estimator proposed in \cite{flatbed} based on a generalized light reflection model to extract norm maps from the preprocessed images. This estimator is capable of modeling both diffuse reflection and potential specular components, thereby providing a more comprehensive representation of the surface characteristics. 
For scanned patches (middle panel, top row of Fig.~\ref{fig:framework}), the estimation of normal vectors requires the acquisition of four images scanned at 0\textdegree, 90\textdegree, 180\textdegree\ and 270\textdegree\ as first proposed in \cite{clarkson2009fingerprinting}. 
If the acquisition is made using mobile devices (middle panel, bottom row of Fig.~\ref{fig:framework}) as in \cite{wong2017counterfeit}, four or more images are required to solve the overdetermined system of linear equations about the three unknown components of each normal vector. 
This stage, responsible for feature extraction, is the most vulnerable to generative attacks (Sec.~\ref{Attacks_Stage_2}) and surrogate modeling (Sec.~\ref{Attacks_Stage_2}), aiming to model the underlying feature distribution.

\smallskip
\noindent
\textbf{Stage 3: Storing Reference Data in $\mathcal{D}$.} The storage database~$\mathcal{D}$, shown in the right panel of the top row of Fig.~\ref{fig:framework}, stores the template norm maps acquired during the registration phase. These references will be retrieved during the verification phase by the decision-making system to assist in the computation of an authentication decision. Since scanners offer a better resolution of the microstructures of the surfaces, the templates can be registered into the database acquired by a scanner in the registration phase. The reference norm map is then stored as an (ID, Template) pair for retrieval purposes. Alternatively, the reference templates can also be stored without an associated ID, and in this case, the verification process will perform a similarity search over the whole $\mathcal{D}$. This stage, responsible for storing reference authentication features, is the most vulnerable to insider attacks (Sec.~\ref{Attacks_Stage_3}), aiming to leak reference data, and reverse engineering attacks (Sec.~\ref{Attacks_Stage_3}), aiming to use leaked data to input synthetic images into the authentication system.

\begin{table*}[!ht]
    \caption{Overview of Practical Threat Models\label{table:overview}}
    \small
{\renewcommand{\arraystretch}{1.15}
    \begin{tabular}{>{\centering\arraybackslash}p{5cm}|>{\centering\arraybackslash}p{5cm}>{\centering\arraybackslash}p{5cm}>{\centering\arraybackslash}p{3cm}}
    \specialrule{0.5pt}{0pt}{0pt}
    \multicolumn{1}{c|}{\textbf{Adversarial Goal}} & \multicolumn{1}{c|}{\textbf{Adversary's Knowledge}} & \multicolumn{1}{c|}{\textbf{Attack Type}}         & \textbf{Attack Action}                                         \\ \hline
    \multicolumn{1}{c|}{\multirow{4}{*}{Sabotaging Authentication}} & \multicolumn{1}{c|}{\multirow{4}{*}{\shortstack[c|]{No paper PUF knowledge, \\Has access to product packaging \\
    E.g., Factory worker}}} & \multicolumn{1}{c|}{\multirow{4}{*}{Physical Denial-of-Service (DoS)}} &  Scratching                                                      \\ \cline{4-4} 
    \multicolumn{1}{c|}{} & \multicolumn{1}{c|}{} & \multicolumn{1}{c|}{}                                 & \begin{tabular}[c|]{@{}c@{}} Physical Patching \end{tabular} \\ \cline{4-4} 
    \multicolumn{1}{c|}{} & \multicolumn{1}{c|}{}  & \multicolumn{1}{c|}{}                                & Scribbling                                                      \\ \cline{4-4} 
    \multicolumn{1}{c|}{} & \multicolumn{1}{c|}{}  & \multicolumn{1}{c|}{}                                & Crumpling                                                       \\ \cline{4-4} 
     \hline
    \multicolumn{1}{c|}{\multirow{2}{*}{Circumventing Authentication}} & \multicolumn{1}{c|}{\multirow{2}{*}{\shortstack[c|]{\shortstack[c|]{Paper PUF expert \\ E.g., Counterfeiting scientist}}}} & \multicolumn{1}{c|}{\multirow{2}{*}{Digital Forgery}} & \multicolumn{1}{c}{\multirow{2}{*}{Hill Climbing}} 
    \\  
    \multicolumn{1}{c|}{} & \multicolumn{1}{c|}{} & \multicolumn{1}{c|}{}                                 & \multicolumn{1}{c}{} \\
    \hline 
    \specialrule{0.5pt}{0pt}{0pt}
    \end{tabular}}
    \vspace{-2mm}
\end{table*}

\smallskip
\noindent
\textbf{Stage 4: Decision-Making Using $\delta$.} The decision-making stage is responsible for generating a binary answer signifying the success or failure of the authentication. As shown in the right panel of the bottom row of Fig.~\ref{fig:framework}, for the $k$th product, this stage computes $\phi(f(x_{\text{q}}))$. It is then compared against $\phi(f(x_k))$ if the product ID is available, or otherwise against $\{\phi(f(x_{i}))\}_{i=1}^N$, where $N$ is the total number of templates in the reference database~$\mathcal{D}$. The comparison may be done using distance metrics such as the $\ell_{2}$ distance or the Pearson correlation, e.g., $\delta(x_{\text{q}},x_{k},\epsilon)=\mathds{1}(\|\phi(f(x_{\text{q}})) - \phi(f(x_k))\|_{2} \leq \epsilon) $ for a chosen similarity threshold $\epsilon$. This stage, responsible for matching the reference and query features and outputting a decision, is the most vulnerable to optimization-based attacks such as hill climbing (Sec.~\ref{Sec4.3}, Sec.~\ref{Attacks_Stage_4}) to digitally forge features.

\vspace{-2.5mm}

\section{Exposing Paper PUF Vulnerabilities\label{Sec4}}

While multiple vulnerabilities may exist in paper-PUF-based authentication systems, practical constraints often limit the feasibility for adversaries to launch attacks. 
We use this section to expose such real-world vulnerabilities that may be exploited and lead to practical attacks, and then expand the discussion in Section~\ref{Sec5} to delineate other possible attacks.

\vspace{-2.5mm}

\subsection{Overview of Practical Adversarial Goals}

To expose practically meaningful vulnerabilities, we first define Mallory's (the adversary's) knowledge and goals, as summarized in Table~\ref{table:overview}.  
First, Mallory may lack system knowledge---such as being a factory worker specializing in counterfeiting but not a security expert.
Such adversaries are common in high-volume e-commerce supply chains, such as fashion goods \cite{adv_goal_1_fashion}, and may also penetrate critical industries, such as currency production \cite{adv_goal_1_money}.
In such situations, Mallory may resort to sabotaging authentication.
Second, Mallory could be a scientist-for-hire with domain expertise in paper-PUF-based authentication, allowing her to develop more sophisticated strategies to circumvent authentication. 
Such adversaries are typical in high-expertise environments, such as electronic component design \cite{adv_goal_2}.
To select appropriate attacks for both adversarial goals, we elaborate on Mallory's objectives. 

\smallskip
\noindent \textbf{Adversarial Goal \#1: Sabotaging Authentication}. For an adversary without domain expertise and limited access to only the product’s packaging, Mallory has few viable options. She may attempt to find paper sheets with microscopic imperfections similar to those in the genuine packaging. However, the authentication error rates of norm-map-based authentication systems are on the order of $10^{-13}$--$10^{-130}$~\cite{wong2017counterfeit}, demonstrating the low likelihood of finding a sufficiently similar sheet of paper. Therefore, the most feasible strategy for such an adversary is to undermine paper-PUF-based authentication systems by launching physical denial-of-service (DoS) attacks on genuine products. These attacks could cause the system to fail within its intended operating conditions, making it impossible to distinguish between physically damaged genuine and counterfeit products, and thereby introducing uncertainty into the verification process. 
Although Toreini et al.~\cite{toreini2017texture} and Clarkson et al.~\cite{clarkson2009fingerprinting} previously studied the physical resilience of paper-PUF estimation against handling issues such as wetting and scribbling, they did not consider attacks by an informed adversary exploiting such wear and tear, while concluding such systems to be robust. Section~\ref{Sec4.2} will experimentally demonstrate the feasibility and effectiveness of various physical denial-of-service (DoS) attacks, including scratching with keys, scribbling with ink, concealing portions of paper surfaces, and crumpling or folding paper to distort it, in sabotaging paper-PUF-based authentication systems. 
Such attacks target the first stage of the operational framework presented in Section~\ref{Sec3} and in Fig.~\ref{fig:framework}.

\smallskip
\noindent \textbf{Adversarial Goal \#2: Circumventing Authentication}. Although physical denial-of-service attacks may introduce ambiguity into the authentication process, Mallory may struggle to circumvent authentication in a targeted manner. Specifically, she may not be able to provide Alice with counterfeit products and have the counterfeit products authenticated as genuine. Achieving this more sophisticated goal requires Mallory to be an informed adversary, utilizing the intricacies of the framework presented in Section~\ref{Sec3}. In particular, Mallory would need domain expertise in paper-PUF-based authentication systems, similar to a scientist working with counterfeiters. 

To successfully disguise a counterfeit product as genuine, she would need to modify the fake product in a way that aligns its PUF features with the reference features stored in Bob’s database. Manufacturing specific microstructures or tampering with paper surfaces to manipulate PUF extraction is a highly challenging task for adversaries, due to the need to alter the complex and random intertwining of cellulose fibers---the very characteristic making paper surfaces physically unclonable. Rather than manufacturing physical forgeries, we explore Mallory's ability to create digital forgeries of the paper surface in Section~\ref{Sec4.3}. 
By penetrating the user client responsible for signal acquisition and feature extraction, and carefully crafting the digital features sent to the reference database, Mallory can potentially have the counterfeit product authenticated as genuine. 
Assuming Mallory possesses security expertise, we propose that she could infiltrate Alice’s client, which extracts these features, and execute optimization attacks to generate synthetic authentication features that mimic genuine ones, allowing the counterfeit product to pass authentication. 
Such attacks target the second, third, and fourth stages of the operational framework presented in Section~\ref{Sec3} and in Fig.~\ref{fig:framework}.
In this work, we focus on demonstrating the feasibility of digital forgeries in circumventing a single authentication step. 
The continual misauthentication of products over time is left for future research, as it requires modeling complex multi-adversarial collaboration.

\vspace{-3mm}

\subsection{Achieving Goal \#1 via Physical DoS Attacks\label{Sec4.2}}

\begin{figure}
        \vspace{-3.5mm}
        \centering
        \subfloat[\label{fig_2c}]{\includegraphics[width=0.45\linewidth]{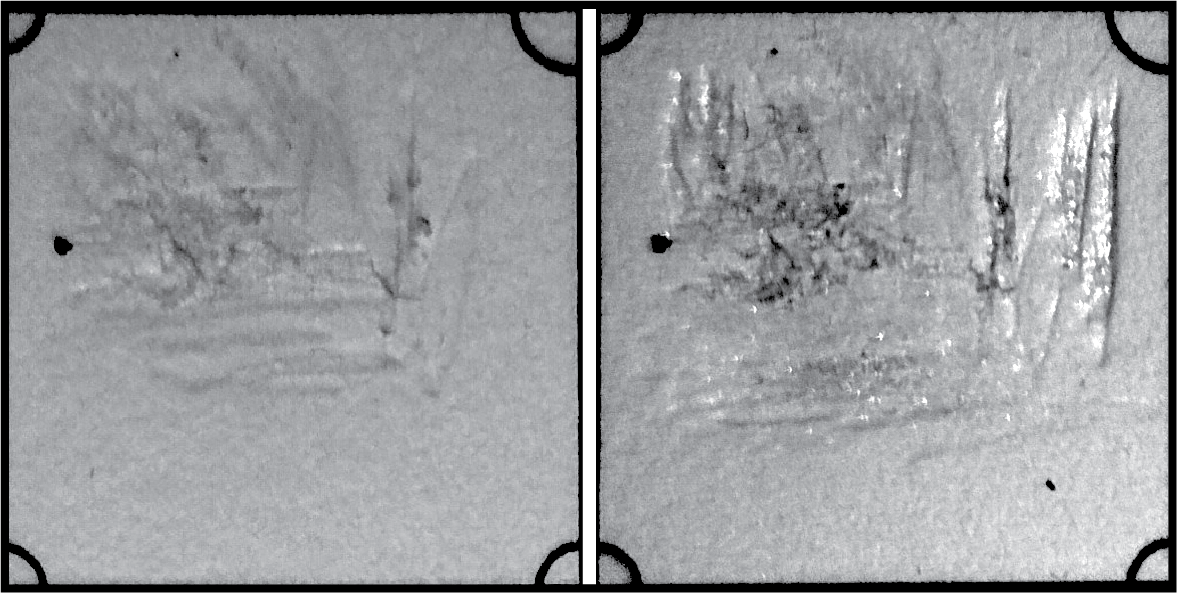}}
        \hfill
        \subfloat[\label{fig_2d}]{\includegraphics[width=0.45\linewidth]{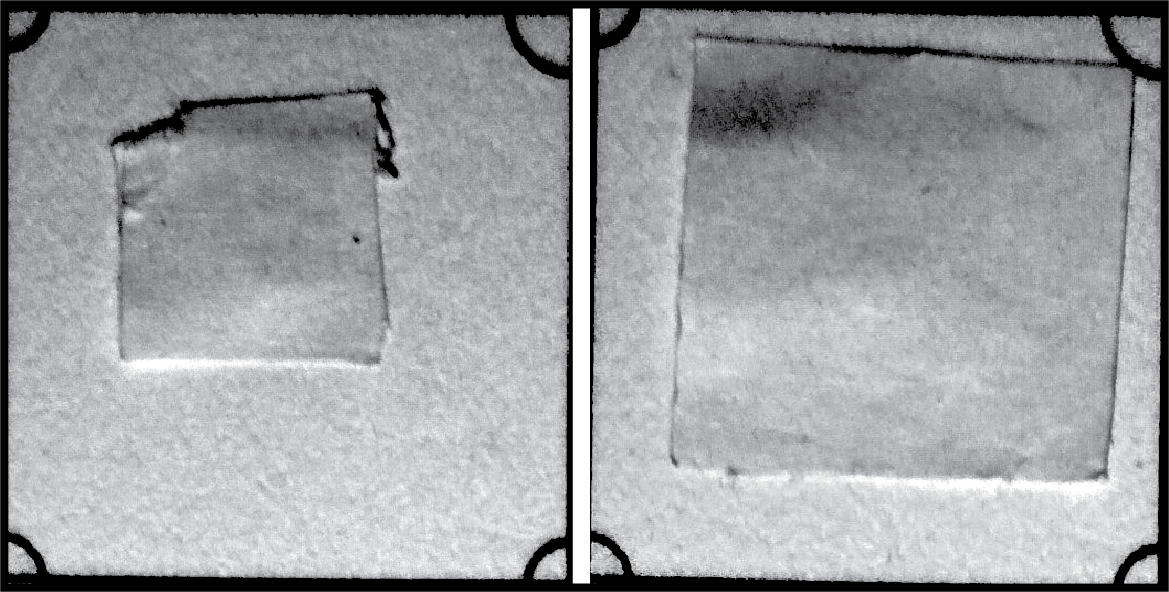}}
        
        \subfloat[\label{fig_2e}]{\includegraphics[width=0.45\linewidth]{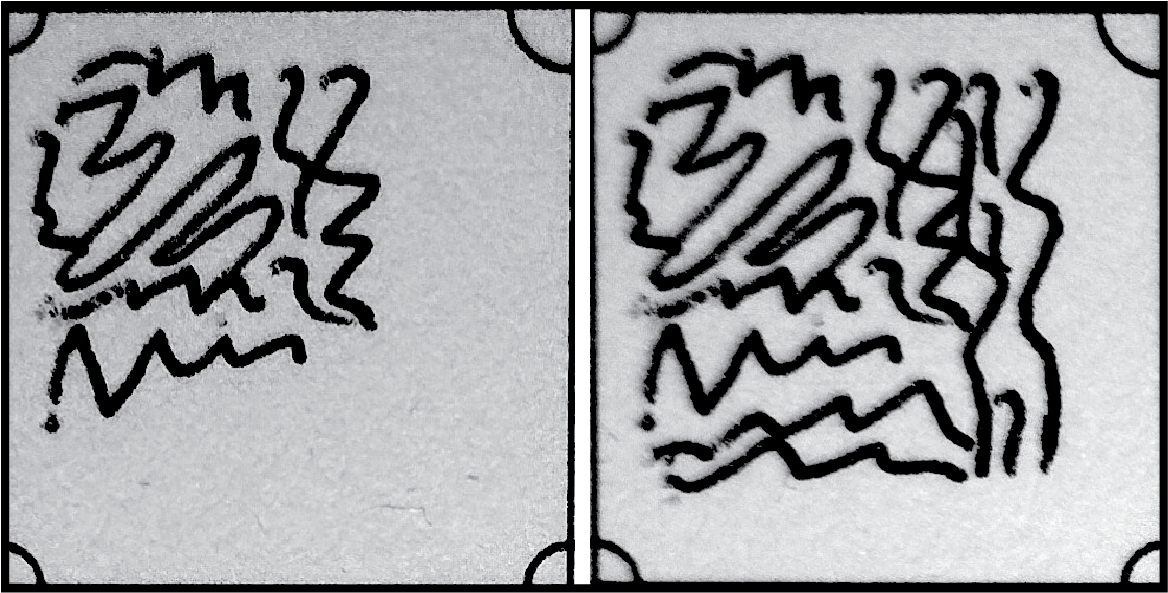}}
        \hfill
        \subfloat[\label{fig_2f}]{\includegraphics[width=0.45\linewidth]{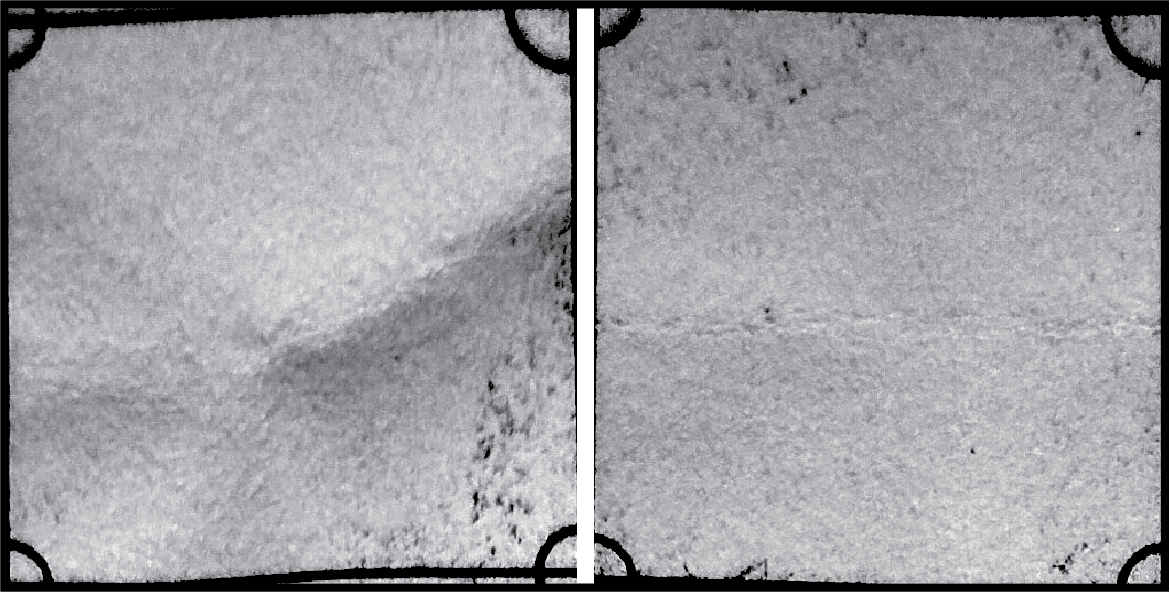}}
    \caption{Geometrically aligned images of paper patches after the following physical attacks: (a)~scratching, (b)~physical patching, (c)~scribbling with 25\% and 50\% attack strengths, respectively, and (d)~crumpling.\label{fig:DoS_Real_Img}}
    \vspace{-4mm}
\end{figure}

In this subsection, we investigate the feasibility of sabotaging authentication by Mallory, an adversary without expertise in paper-PUF-based authentication systems. We define a precise threat model for launching physical DoS attacks, describe the experimental setup,
and analyze their effectiveness. We launch four practical attacks: scratching, physical patching, scribbling, and crumpling, visualized in Fig.~\ref{fig:DoS_Real_Img}(a)--(d) respectively.

\smallskip
\noindent
\textbf{Threat Model.} The paper-PUF-based authentication system is assumed to be deployed for large-scale supply chains, where each product with paper packaging passes through multiple stakeholders. 
Mallory, a stakeholder with access to the genuine product's packaging, aims to disrupt proper authentication through damaging the paper packaging.
By launching physical DoS attacks on a significant number of products in the supply chain, Mallory can achieve several outcomes. Damaging numerous PUFs introduces confusion into the authentication system, potentially causing it to fail, which could lead to the suspension or bypassing of PUF-based authentication. Additionally, subtle tampering of genuine packaging---such as scratching, folding, or ink smudging---could be mistaken for normal wear and tear during transit, avoiding detection. This may prevent the system from distinguishing between authentic and counterfeit products, facilitating the introduction of additional counterfeit items into the supply chain. Unlike traditional anti-counterfeiting solutions like QR codes, paper-PUF-based systems lack error correction capabilities, making their correlation mechanisms vulnerable to such straightforward attacks.
We have considered four types of physical attacks.

\begin{enumerate}[left=0pt]
    \itemsep=3pt 
    \item \underline{Scratching:} A selected portion of a paper surface is intentionally scratched with a metal key. The scratched area is gradually increased from 5\% to 10\%, 25\%, 50\%, and 75\%, and in each case, the correlation coefficient of the extracted norm maps with the reference norm maps is calculated.
    \item \underline{Physical Patching:} We cover up a portion of the paper patch with sticker paper and then take photos of the paper surface to perform authentication. The attack is conducted at the same strengths as the scratching attack (5\%, 10\%, 25\%, 50\%, and 75\% of the surface area).
    \item \underline{Scribbling:} We randomly scribble a portion of the paper patch surface using a ballpoint pen. We have chosen ballpoint pens due to their common availability. We do not consider gel pens because variations in the diffusive characteristics of their ink may affect the authentication performance differently. Similar to scratching and physical patching, this attack is also conducted at 5 different attack strengths.
    \item \underline{Crumpling:} We consider two cases. First, we randomly crumple a paper patch to the extent that visible distortion can be noticed on the paper patch. Second, we fold a paper patch twice in two directions. A point inside the patch is folded along the $y$-axis and then it is folded along the $x$-axis. For both attacks, we smooth out the attacked paper surface using a steam iron to mimic an honest user's effort to ensure proper authentication.
\end{enumerate} 

\begin{figure}[!t]
    \vspace{-4mm}
    \centering
    \subfloat[\label{fig_3a}]{\includegraphics[width=0.5\linewidth]{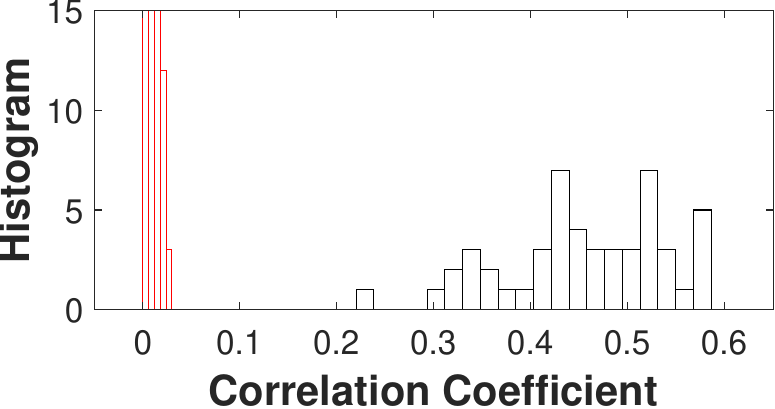}}  
    \subfloat[\label{fig_3b}]{\includegraphics[width=0.5\linewidth]{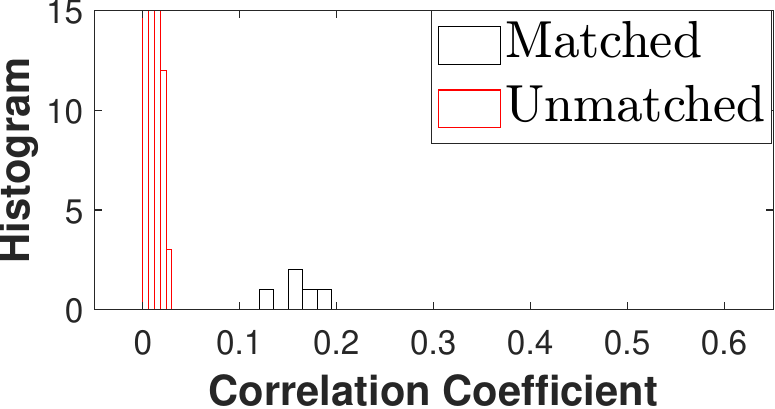}}

    \subfloat[\label{fig_3c}]{\includegraphics[width=0.5\linewidth]{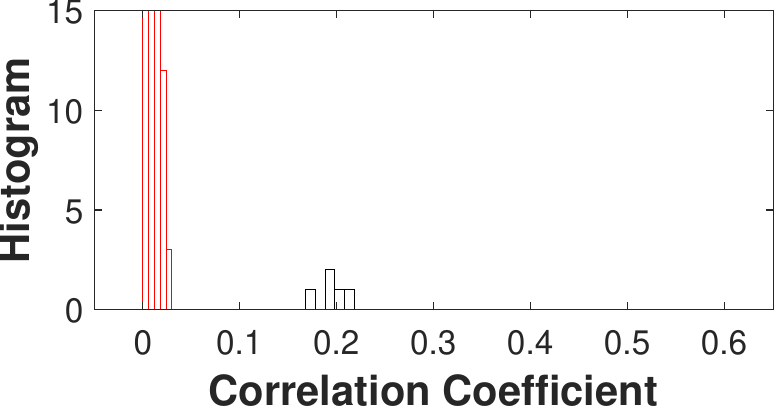}}
    \subfloat[\label{fig_3d}]{\includegraphics[width=0.5\linewidth]{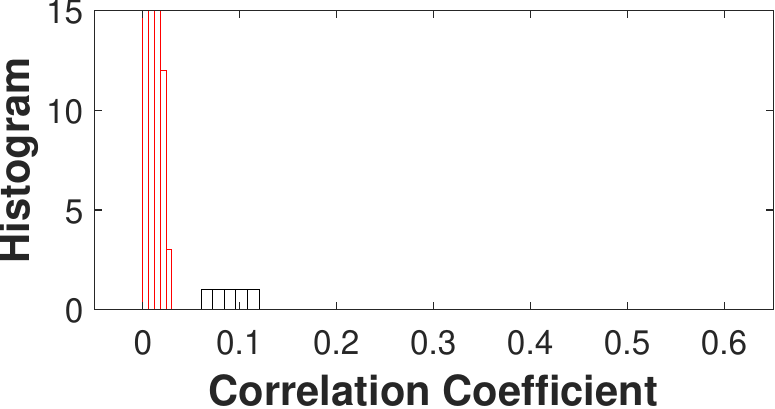}}
    \caption{\label{fig:histogram_DoS}Impact of physical denial-of-service attacks to correlation under matched (in black) and unmatched cases (in red) (a)~without attack, and with (b)~scratching attack, (c)~physical patching attack, and (d)~scribbling attack. All three physical DoS attacks (performed at the 25\% attack strength) successfully narrow the gaps between the histograms of matched and unmatched cases, thereby reducing authentication accuracy.}
    \label{dist}
    \vspace{-4mm}
\end{figure}

\smallskip
\noindent
\textbf{Experimental Conditions and Baseline.} We replicate the paper-PUF-based authentication system from \cite{wong2017counterfeit}, which uses mobile cameras to extract norm maps. We extracted norm maps from ten distinct resume paper sheets. We collected five samples from each sheet, before and after the sheet was attacked. For the four types of attacks we examined, each attack was performed on two sheets. We used a Canon 4400f scanner to extract reference norm maps and the Redmi Note 11's primary camera for estimating test norm maps. We visualize the baseline results before the attacks using histograms of sample correlations for the matched and unmatched cases in Fig.~\ref{dist}(a). The rest of the histograms in Fig.~\ref{dist} visualize the first three types of attack at 25\% attack strength. 
For these histograms, greater gaps between matched and unmatched cases indicate better authentication performance.   

\smallskip
\noindent
\textbf{Results.}
Table~\ref{dos_result} presents the statistics of correlations for matched cases across all attacks at three attack strengths. 
Comprehensive results across all five attack strengths are provided in Appendix~\ref{App:A_DoS} of the supplemental material.
As the attack strength increases for any of the four physical attacks, the average correlation drops, indicating an inverse relation between the correlation of matched cases and the portion of paper attacked. 
For scratching and physical patching, the average $x$ correlation drops from 0.46 before the attack to $\sim$0.2 when attacked at 25\% strength. 
These post-attack matched correlations are still separable from the unmatched correlations. 
When the attack strength increases further, the matched correlations drop into the unmatched region.

\begin{table}[!ht]
\centering
\caption{Effect of Physical DoS Attacks at 5\%, 25\%, and 75\% Strengths on Matched Case Correlation.}
\label{dos_result}
{\renewcommand{\arraystretch}{1.3}
\begin{tabular}{cccc}
\toprule
\textbf{\begin{tabular}[c]{@{}c@{}}Physical \\ Attack\end{tabular}} & \textbf{\begin{tabular}[c]{@{}c@{}}Attack \\ Strength (\%)\end{tabular}} & \textbf{\begin{tabular}[c]{@{}c@{}}$\boldsymbol{x}$ Corr Coef\\ Mean $\downarrow$ (Std)\end{tabular}} & \textbf{\begin{tabular}[c]{@{}c@{}}$\boldsymbol{y}$ Corr Coef\\ Mean $\downarrow$ (Std)\end{tabular}} \\ \hline
\textbf{No Attack} & N/A & 0.46 (0.08) & 0.36 (0.09) \\ \hline 
\multirow{3}{*}{\textbf{Scratching}} & 5 & 0.37 (0.02) & 0.16 (0.07) \\ \cline{2-4} 
 
 & 25 & 0.16 (0.02) & 0.05 (0.04) \\ \cline{2-4} 
 
 & 75 & 0.03 (0.01) & 0.01 (0.01) \\ \hline
\multirow{3}{*}{\textbf{\begin{tabular}[c]{@{}c@{}} Physical \\ Patching\end{tabular}}} & 5 & 0.45 (0.03) & 0.17 (0.04) \\ \cline{2-4} 
 
 & 25 & 0.20 (0.02) & 0.09 (0.03) \\ \cline{2-4} 
 
 & 75 & 0.03 (0.00) & 0.01 (0.01) \\ \hline
\multirow{3}{*}{\textbf{Scribbling}} & 5 & 0.20 (0.02) & 0.13 (0.03) \\ \cline{2-4} 
 
 & 25 & 0.09 (0.02) & 0.02 (0.01) \\ \cline{2-4} 
 
 & 75 & 0.01 (0.01) & 0.01 (0.01) \\ \hline
\multirow{2}{*}{\textbf{Crumpling}} & \begin{tabular}[c]{@{}c@{}}Random\end{tabular} & 0.03 (0.02) & 0.06 (0.03) \\ \cline{2-4} 
 & \multicolumn{1}{c}{Folding} & 0.02 (0.01) & 0.08 (0.02) \\ 
\bottomrule
\multicolumn{4}{l}{``$\downarrow$'' indicates that a lower mean is desired by attackers aiming to} \\
\multicolumn{4}{l}{sabotage authentication.}
\end{tabular}}
\vspace{-4mm}
\end{table}

Scribbling attacks exhibit a similar downward trend but at just 25\% attack strength, the matched $x$ correlation of 0.09 is firmly in the unmatched region. 
This observation indicates that scribbling using black ink on paper affects the surface normal estimation more severely than either scratching or physical patching.
For both random crumpling and folding attacks, the matched $x$ correlation is almost negligible (0.02--0.03) because these attacks affect the image alignment process, an important factor in paper-PUF-based authentication \cite{wongicip}.
The $y$ correlations exhibit similar trends to the $x$ correlations. 
Based on the experimental evidence, it is apparent that the various physical DoS attacks are effective in disrupting paper-PUF-based authentication systems, which were previously 
considered resilient to such tampering \cite{clarkson2009fingerprinting}.

\vspace{-2.5mm}

\subsection{Achieving Goal \#2 via Digital Forgery Attacks\label{Sec4.3}}

In this subsection, we explore the capabilities of Mallory, a sophisticated adversary with domain expertise, to maliciously authenticate a counterfeit product as genuine. 
We design several fast digital attacks that can guess the correct norm map, circumventing the need for the presence of the authentic physical paper patch in the verification process.

\smallskip
\noindent
\textbf{Threat Model.} We assume a large-scale paper-PUF-based authentication system illustrated in Fig.~\ref{fig:framework} follows the client--server model described in \cite{flatbed}. An open-source client encompassing stages~1 and~2 is fully observable and vulnerable to tampering, whereas the server's implementation encompassing stages~3 and~4 remains unknown to the client. The client receives the server’s outputs, including similarity scores calculated between test and reference norm maps. The adversary is an expert on paper-PUF-based authentication systems. Operating on the client side, the adversary must generate a synthetic norm map $\n_{\text{syn}}$ such that the observed similarity score $\rho$ exceeds the authentication threshold $\tau$. The adversary has access to a holdout dataset $\mathcal{D}_{\text{hold}}$ comprising norm maps derived from scanning a collection of \textit{distinct paper sheets} manufactured using the same material type as the packaging substrate of the genuine product.  
Such an approach is similar to the use of development sets in optimization-base biometric spoofing attacks, commonly referred to as ``hill-climbing'', where attackers leverage accessible sensors and materials to approximate target systems \cite{jain1996introduction,jain2006biometrics, maiorana2014hill}. 
Unlike well-studied biometric traits such as fingerprints or faces---where domain knowledge, feature models, and attack methodologies are well established---norm maps represent an emerging modality with little prior work or shared structure to exploit. 
This makes hill-climbing particularly difficult on norm maps due to their high-dimensional, image-like nature. 
To our knowledge, this is the first work to adapt and extend hill-climbing attacks to the norm map domain, highlighting their effectiveness in circumventing paper-PUF-based authentication.
To overcome these challenges, we develop specialized optimization strategies that enable adversaries to generate effective digital forgeries.

\begin{figure}[!t]
    \vspace{-4mm}
    \setlength{\tabcolsep}{18pt}
    \begin{tabular}[c]{@{}cc@{}}
    \centering
        \subfloat[\label{fig:init}]{
       \includegraphics[width=0.375\linewidth]{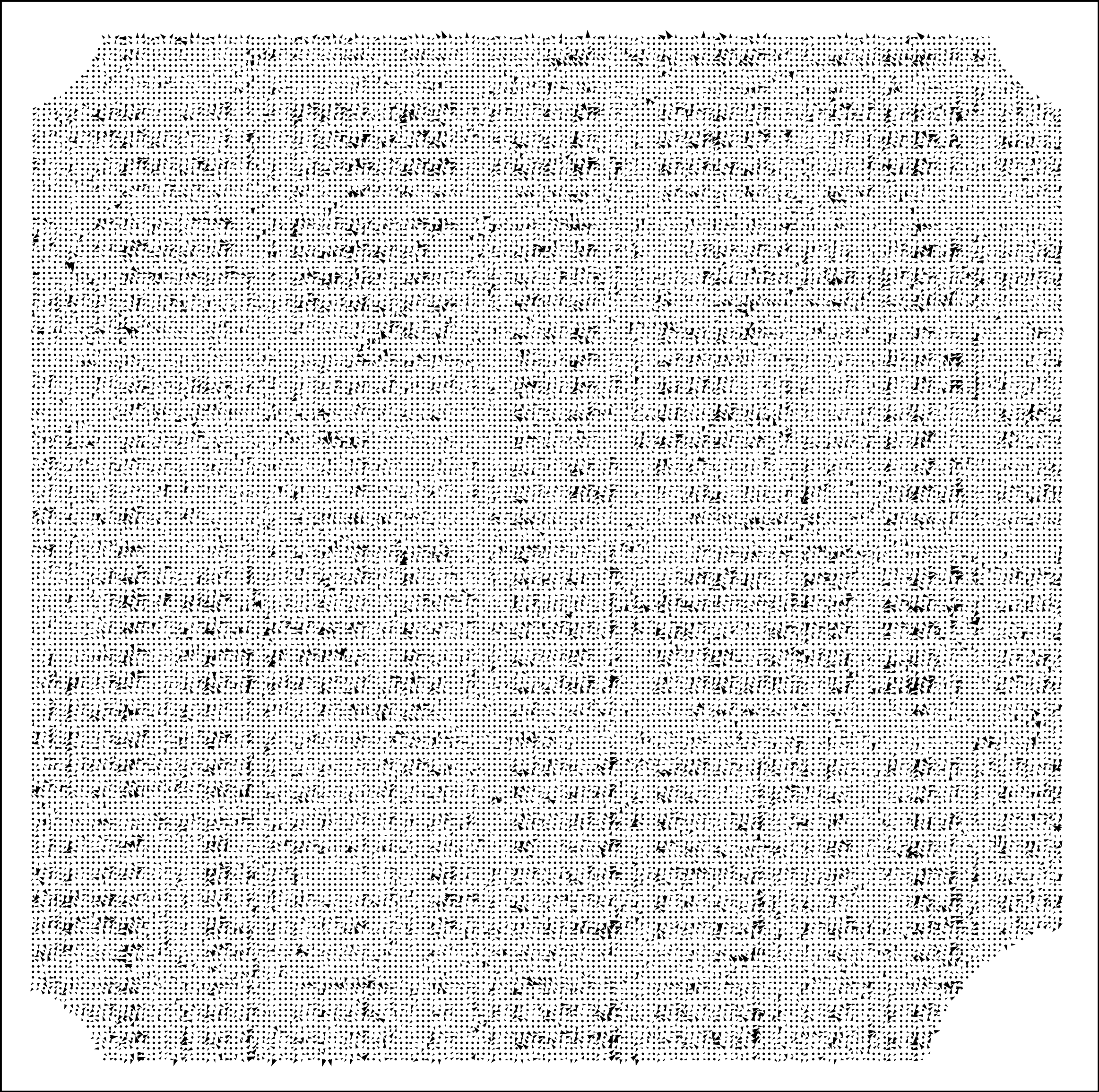}} &
        \subfloat[\label{fig:target}]{
       \includegraphics[width=0.375\linewidth]{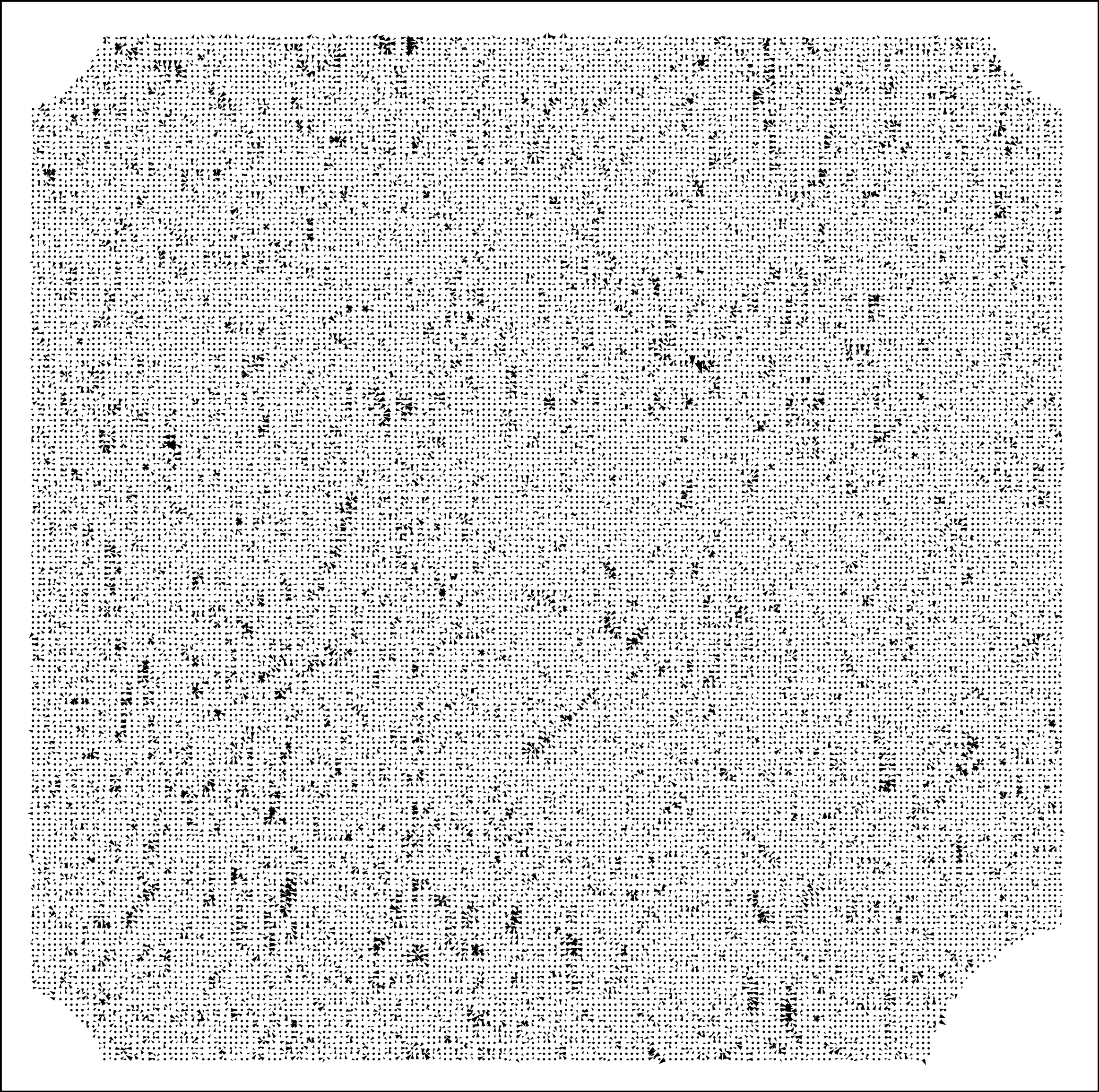}}\\
        \subfloat[\label{fig:synthetic}]{
       \includegraphics[width=0.375\linewidth]{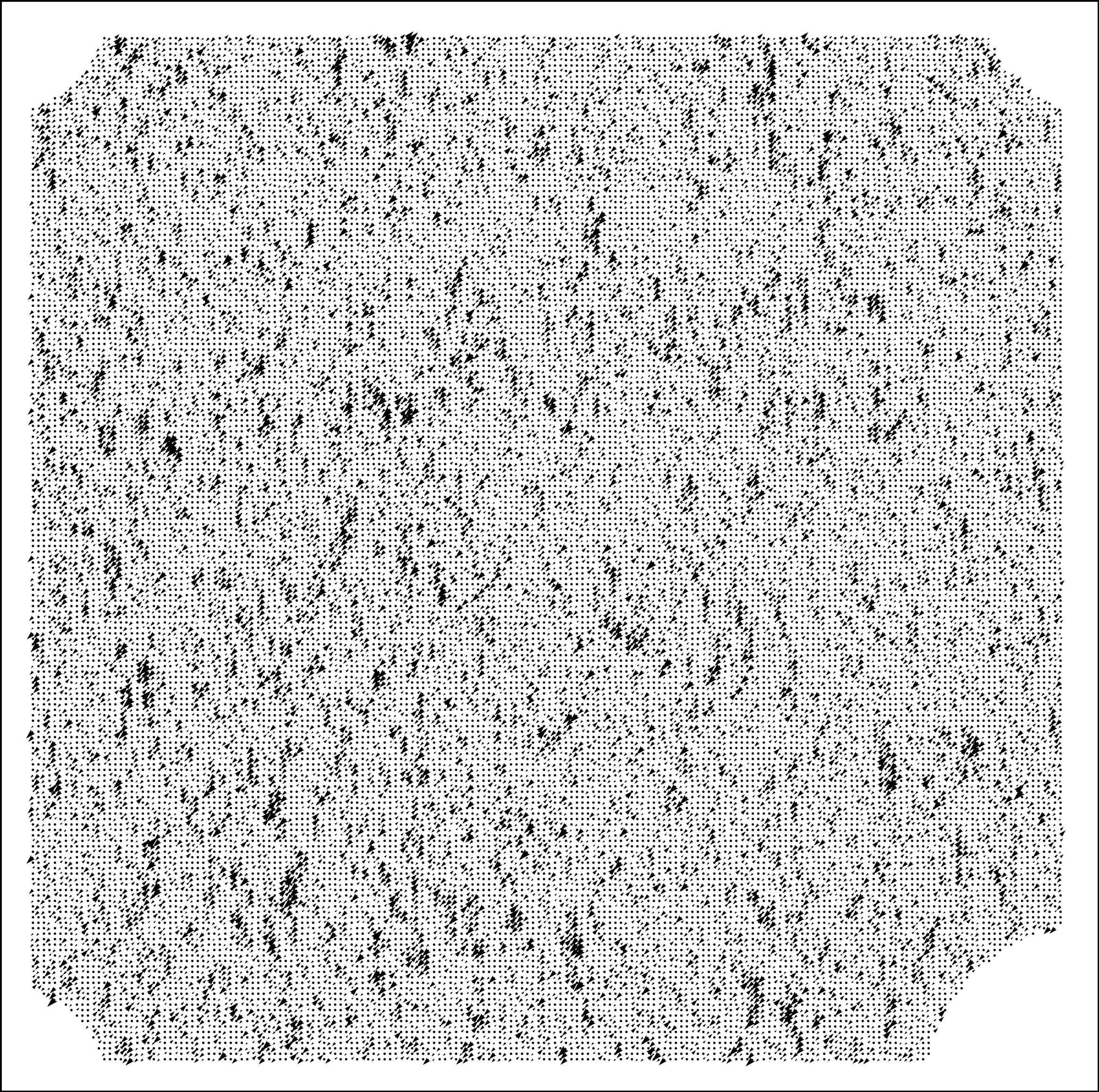}} &
        \subfloat[\label{fig:cosine}]{\includegraphics[width=0.425\linewidth]{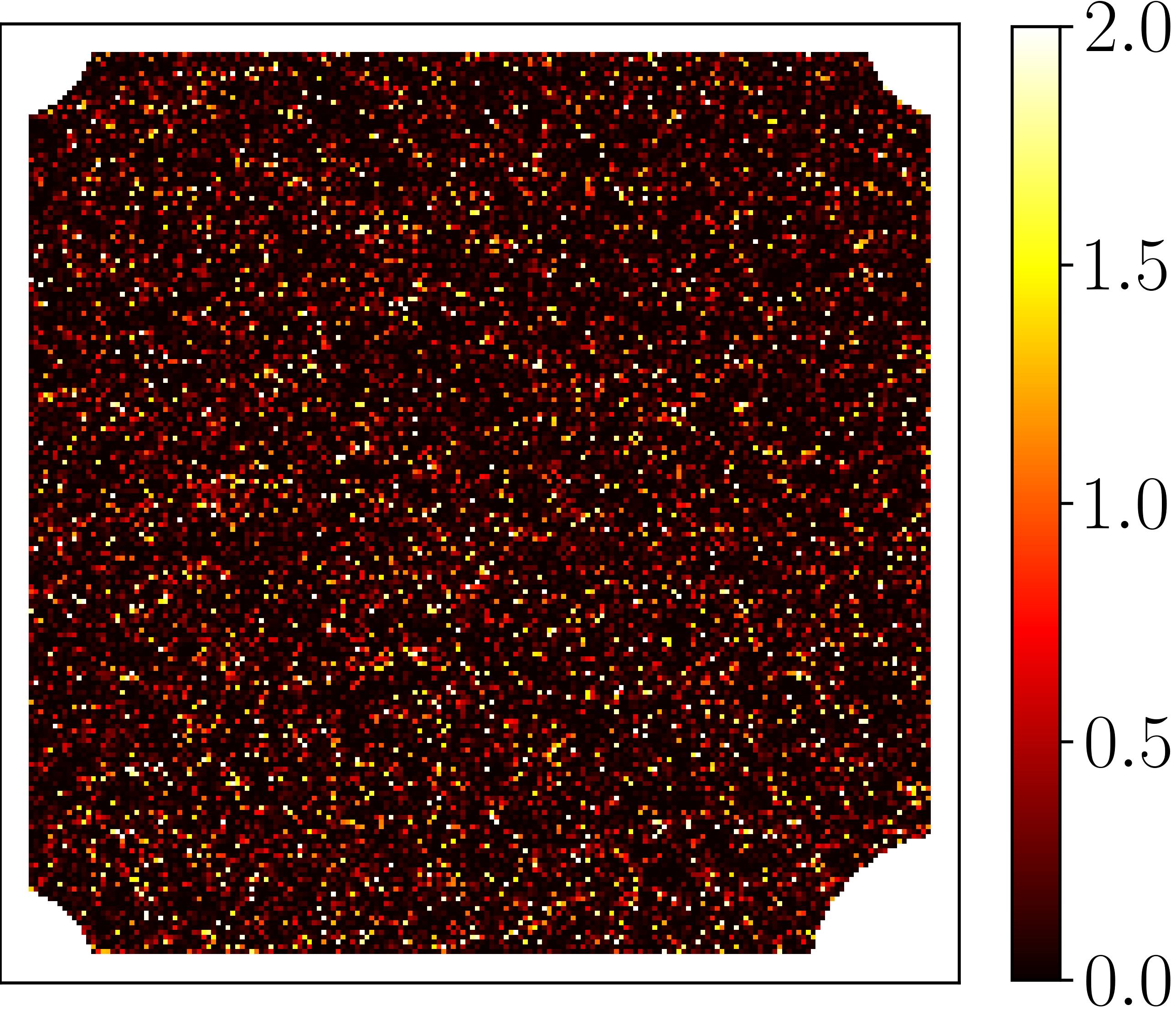}}\\
    \end{tabular}
    \caption{Overview of digital forgery attack: (a)~Initialization achieving a negligibly small correlation of $0.01$, (b)~reference/target norm map, (c)~synthetic norm map generated by a baseline attack w/ $2\times10^5$ function evaluations, achieving a high correlation of $0.91$, and (d)~per-pixel cosine distance between (b) and (c). The digital forgery attack successfully learns the target norm map.}
    \label{fig:overview_forgery}
    \vspace{-3mm}
\end{figure}

\smallskip
\noindent
\textbf{Proposed Digital Attack.} The typical false alarm/positive rates of paper-PUF-based authentication systems range from $10^{-13}$--$10^{-130}$ \cite{clarkson2009fingerprinting,wong2017counterfeit}. This implies that a brute force attack would require at least $10^{13}$ iterations and holdout samples on average, which is impractical. We establish a more realistic high-dimensional baseline attack using a stochastic greedy technique. Starting with a randomly initialized $\n_{0}$, we iteratively refine it. At each iteration $t$, a subset of dimensions chosen randomly from the previous guess, $\n_{t-1}$, is perturbed by noise sampled uniformly from $[-\delta, \delta]$, forming $\n_{t}$. The similarity score $\rho_{t}$ between $\n_{t}$ and the reference norm map is observed. If $\rho_{t}$ exceeds the previous score $\rho_{t-1}$, $\n_{t}$ is accepted as the current best guess; otherwise, $\n_{t}$ is set as $\n_{t-1}$. This process is repeated until iteration $T$, where $\rho_{T} \geq \tau$, to learn a synthetic norm map. However, norm maps are high-dimensional, and a stochastic greedy attack may struggle to explore the vast search space, getting stuck in local minima.

\begin{figure}[!t] 
            \vspace{-3mm}
            \centering
        \subfloat[\label{fig:40hc}]{
       \includegraphics[width=0.5\linewidth]{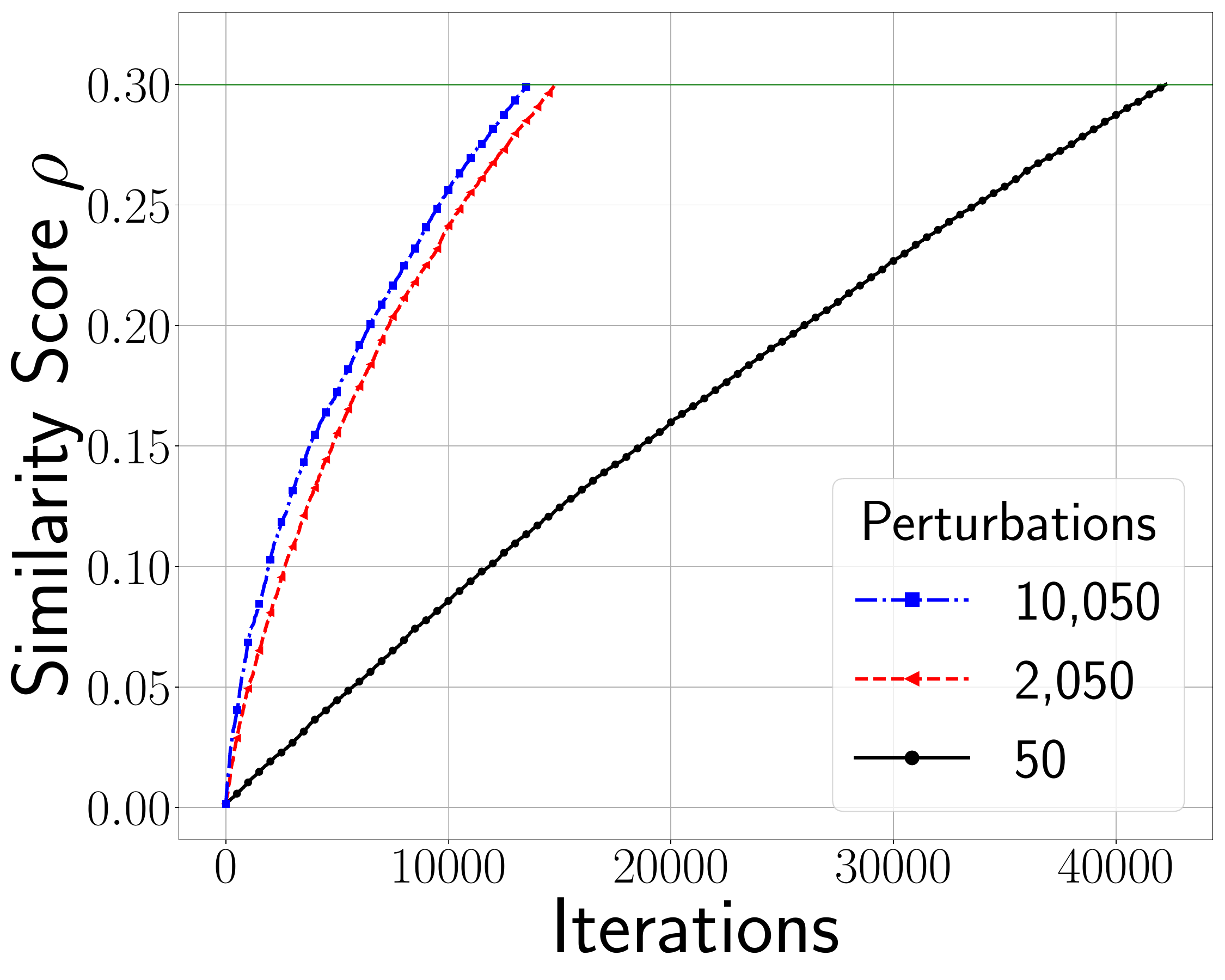}}%
        \subfloat[\label{fig:pcahc}]{
       \includegraphics[width=0.5\linewidth]{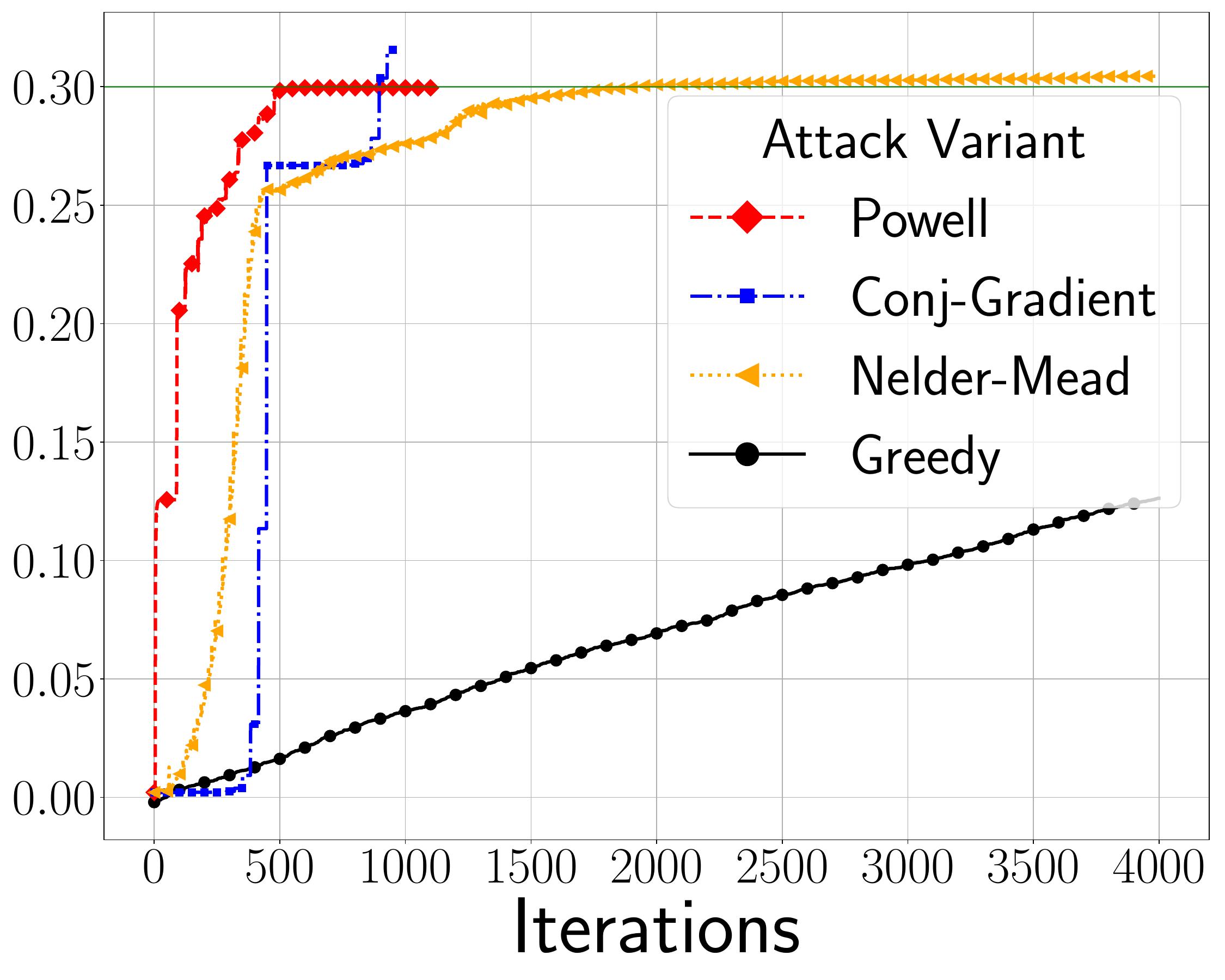}}%
  \caption{(a)~Baseline digital attack as more dimensions are perturbed and (b)~variants of more efficient digital attacks on Sheet 1 in $\mathcal{D}$. For authentication threshold $\tau=0.3$ (in green), all efficient digital attacks succeed within 2000 iterations, $5\times$ faster than the fastest baseline attack in (a).}
  \label{fig:baseline} 
  \vspace{-3mm}
\end{figure}

By leveraging the a priori information $\mathcal{D}_{\text{hold}}$, we examine more efficient attacks that can identify the key linear subspace
to be perturbed. We learn a transformation $h$ on $\mathcal{D}_{\text{hold}}$ that compresses the norm maps into a low-dimensional representation, achievable with techniques such as principal component analysis~(PCA) \cite{pearson1901liii} and variational autoencoders \cite{kingma2013auto}. This compression drastically reduces the search space for the stochastic greedy attack. The attack can then proceed similarly to the baseline attack, except that perturbations are applied in the compressed latent space rather than the feature space. Compression also enables the use of fast black-box optimization techniques such as the Nelder--Mead~\cite{nelder1965simplex} and Modified-Powell~\cite{powell1964efficient} algorithms.
We provide the pseudocode for both baseline and efficient digital attacks in Appendix B of the supplemental document.

\begin{figure*}[!t] 
    \vspace{-4mm}
        \centering
            \subfloat[\label{fig:xy}]{%
            \includegraphics[width=0.28\linewidth]{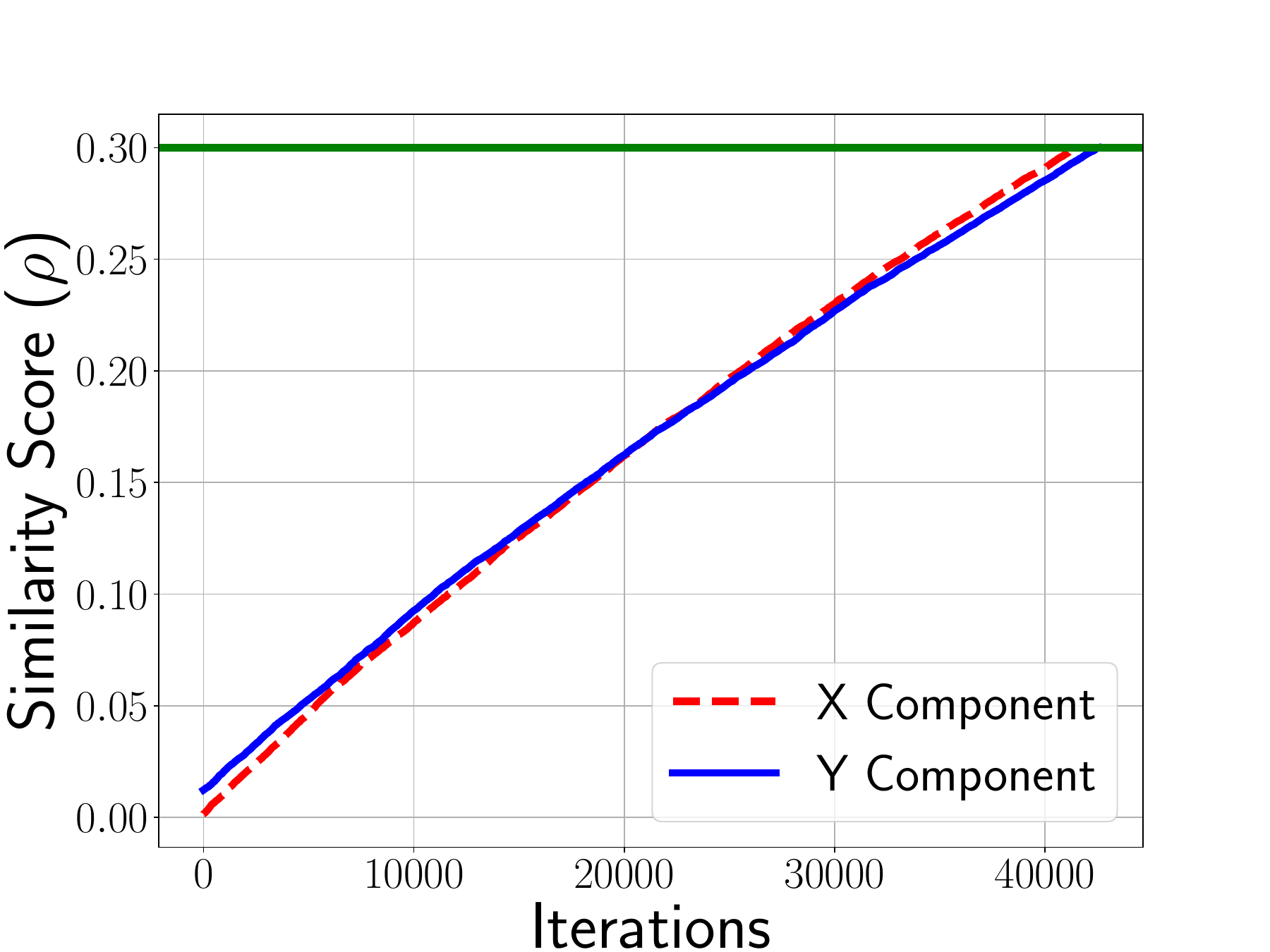}}
            \subfloat[\label{fig:pca}]{%
            \includegraphics[width=0.28\linewidth]{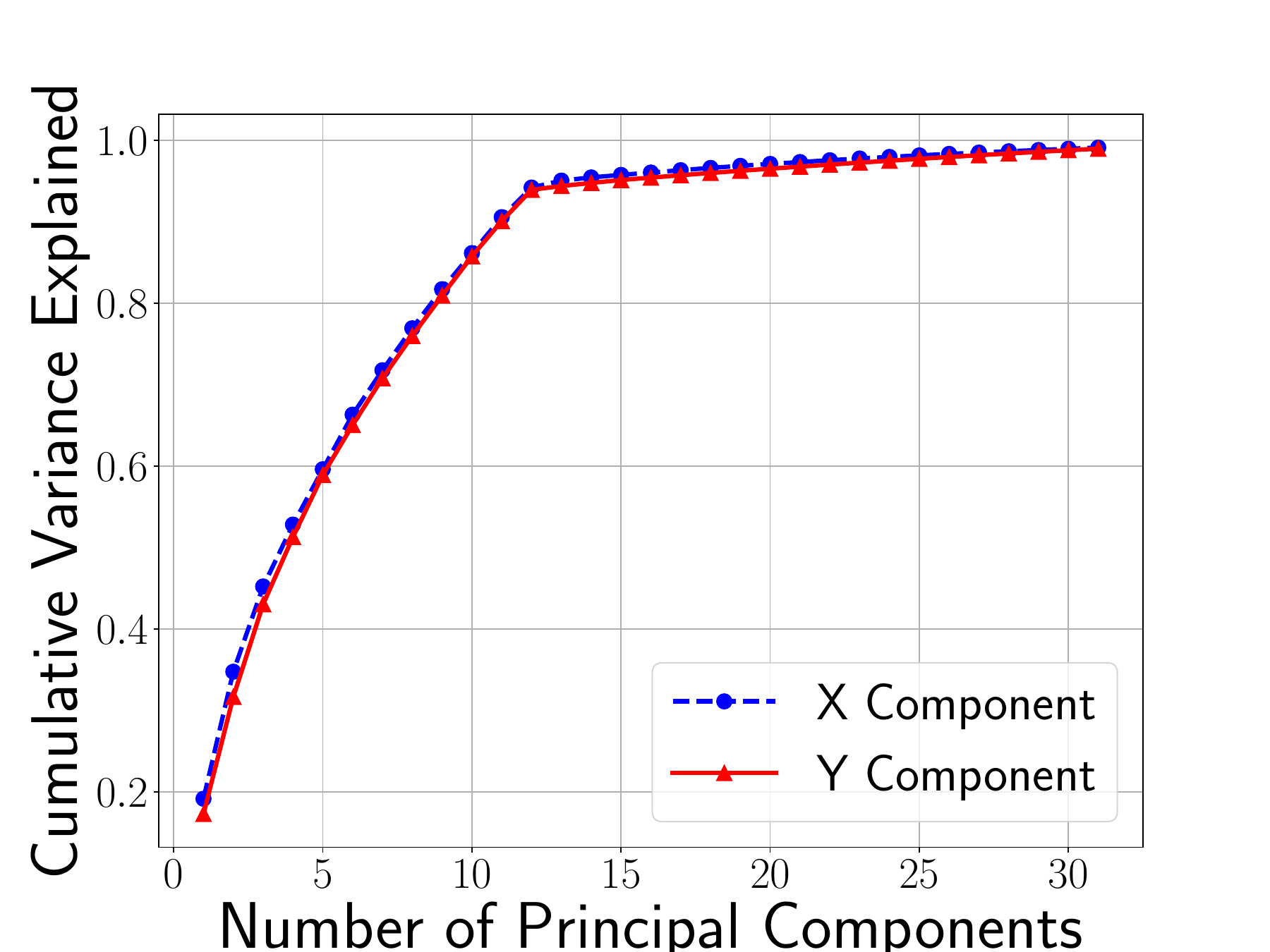}}
            \subfloat[]{
       \includegraphics[width=0.21\linewidth]{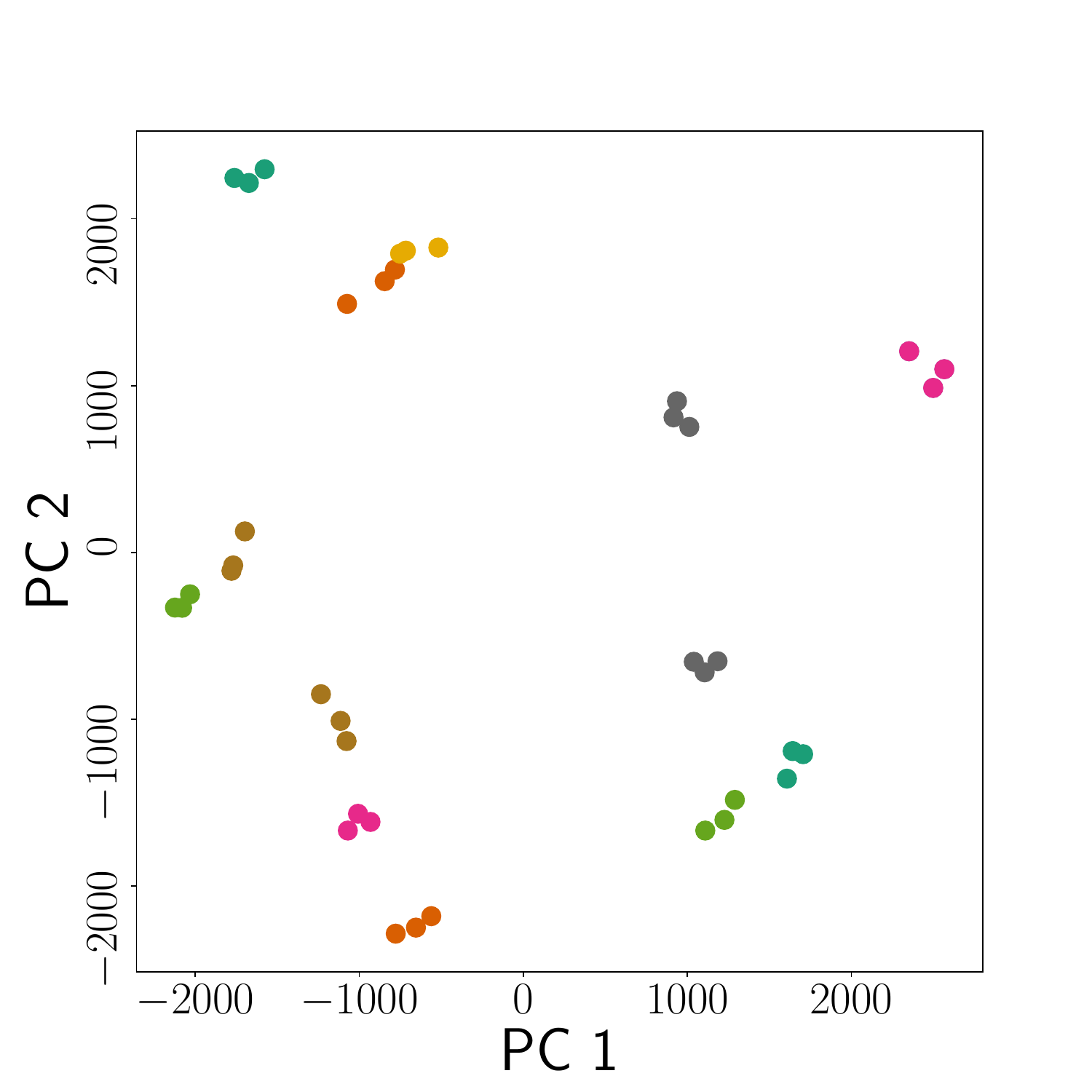}}%
        \subfloat[]{
       \includegraphics[width=0.21\linewidth]{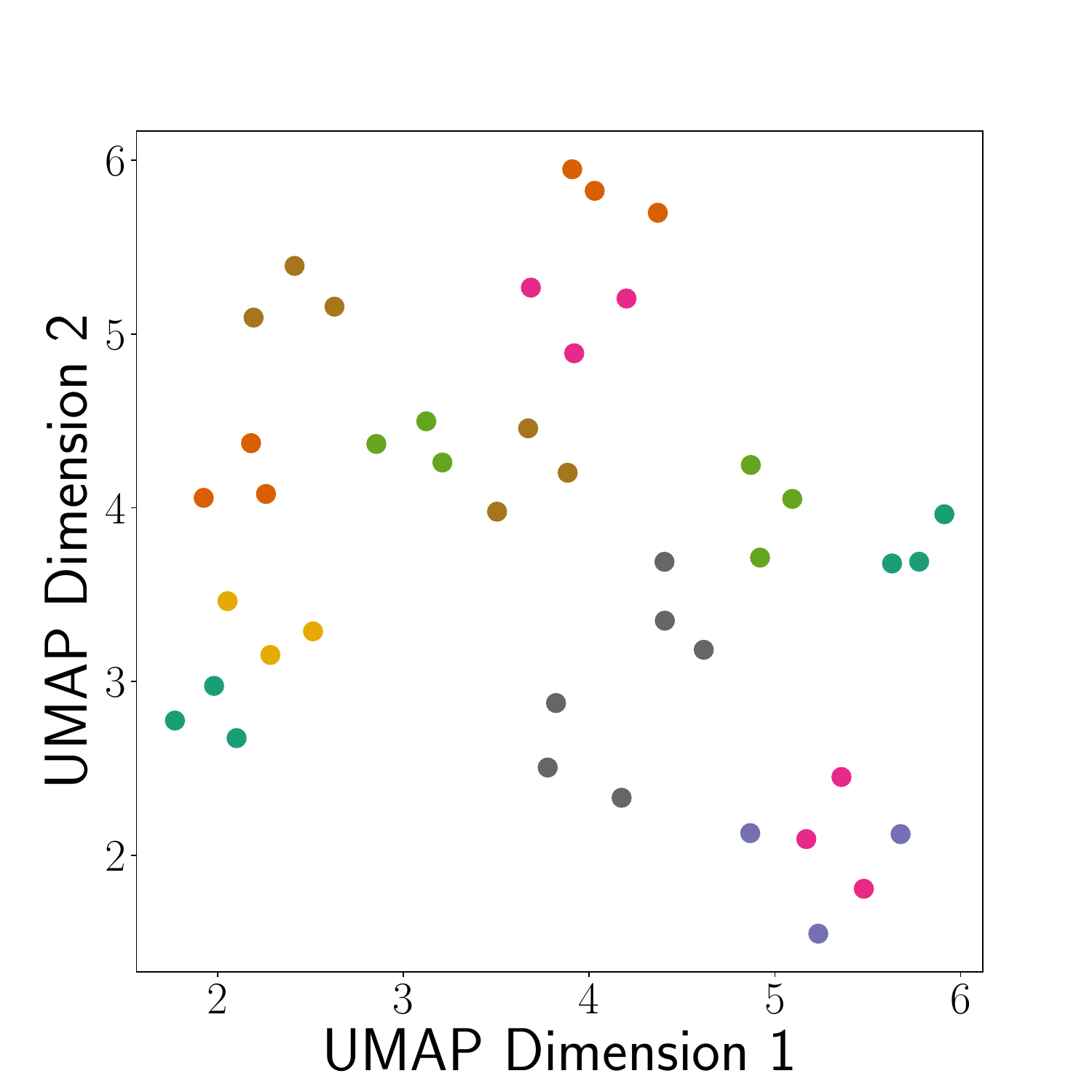}}%
   
  \caption{\label{fig:dimred} (a)~The digital attacks on the $x$- and $y$-components of the reference norm map successfully reach the authentication threshold (in green) in a similar number of iterations. (b)~Using principal component analysis (PCA), 31 principal components capture 99\% of the cumulative variance of norm maps, indicating compressibility. (c)~PCA-compressed PUF representation and (d)~UMAP-compressed PUF representation of Mallory's dataset, with PUFs extracted from the same paper patch shown in the same color. 
  Notably, the differences between the intra-patch spread and inter-patch distance in PCA and UMAP embeddings highlight the crucial role of dimensionality reduction techniques in the effectiveness of efficient digital attack design.   
  This figure is best viewed in color.}
  \label{fig:Efficient} 
  \vspace{-3mm}
\end{figure*}

\smallskip
\noindent
\textbf{Experimental Conditions.} We used 18 distinct resume paper sheets, scanning each sheet three times to create a dataset of 54 samples. 14 sheets were used to construct the adversary's holdout dataset $\mathcal{D}_{\text{hold}}$, whereas the remaining four sheets form the reference database $\mathcal{D}$. We used the scanner-based method from Clarkson et al. \cite{clarkson2009fingerprinting} to extract 200-by-200 pixel norm maps for all samples. We applied PCA \cite{pearson1901liii} to $\mathcal{D}_{\text{hold}}$ to reduce the norm map's dimensionality from 40,000 to 31---the number of dimensions explaining 99\% of the cumulative variance as illustrated in Fig.~\ref{fig:dimred}(b). Based on the separability between histograms of matched and unmatched cases observed in Fig.~\ref{fig:histogram_DoS} and in prior work~\cite{wong2017counterfeit}, $\tau$ was set at 0.3. We used 11,742 iterations---the number of iterations needed by the fastest high-dimensional attack---as a baseline to evaluate the attack success rate of the more efficient attacks across the four reference sheets in Table~\ref{DigAttTable}. 

\smallskip
\noindent
\textbf{Results and Discussions.} While implementing the baseline attack, we empirically observed that $10^4$--$10^5$ function evaluations were sufficient to generate a synthetic norm map, as shown in Fig.~\ref{fig:baseline}(a). 
As shown in Fig.~\ref{fig:dimred}(a), we found that the $x$- and $y$-components of norm maps require similar numbers of iterations to synthesize.
Furthermore, with enough iterations, the baseline attack successfully learned synthetic norm maps closely resembling the target, as shown in Fig.~\ref{fig:overview_forgery}.

\begin{table}[!ht]
\centering
\caption{Attack Success Rates for Digital Attacks.}
\label{DigAttTable}
{\renewcommand{\arraystretch}{1.3}
\resizebox{\linewidth}{!}{%
\begin{tabular}{cccc}
\toprule
\textbf{Feature} & \textbf{Algorithm} & \textbf{Iterations $(\downarrow)$ } & \textbf{Success Rate $(\uparrow)$} \\ \hline
\multirow{2}{*}{\begin{tabular}[c]{@{}c@{}}Raw Norm Map \\ (40k-d)\end{tabular}}
& Brute-Force & $10^{13}$ & N/A \\ \cline{2-4}
& Greedy Baseline & 11,742 & N/A \\ \hline
\multirow{4}{*}{\begin{tabular}[c]{@{}c@{}}PCA Compressed \\ Norm Map \\ (31-d)\end{tabular}}
& Greedy & 5,957 & 0.50 \\ \cline{2-4} 
& Powell & 22 & 0.75 \\ \cline{2-4} 
& Nelder--Mead & 330 & 0.92\\ \cline{2-4} 
& Conjugate-Gradient & 416 & 1.00 \\ 
\bottomrule
\multicolumn{4}{l}{Iterations $(\downarrow)$ and Success Rate $(\uparrow)$ indicate that attackers aiming to circumvent} \\
\multicolumn{4}{l}{authentication prefer faster, more successful attacks.}
\end{tabular}}}
\vspace{-3mm}
\end{table}

The efficient stochastic greedy attack succeeded in approximately $10^3$--$10^4$ evaluations, outperforming the baseline by one order of magnitude. In Table~\ref{DigAttTable}, we report the minimum number of iterations required for each attack variant to reach the authentication threshold. Although difficult to solve in high dimensions, we found that advanced random search optimizers could often find a synthetic norm map in merely $10^1$--$10^2$ evaluations. They also converged much faster than the naive efficient stochastic greedy attacks, as shown in Fig.~\ref{fig:baseline}(b). 
In realistic deployment scenarios, such as factory settings where client devices (e.g., a flatbed scanner) are responsible for authenticating large batches of products, isolating the authentication attempts by adversaries at the individual product level may not be feasible due to the sheer scale of operation.
Corporations may choose to log and monitor clients with preassigned IP addresses, but multiple malicious end-users may choose to collude to send authentication attempts tactically to bypass this restriction.
This lack of granularity in client-level attribution is significant as supply chains process millions of products rapidly, masking $10^2$ failed queries.
Additionally, using our conservative baseline of $10^4$ iterations, advanced black-box optimizers successfully compromise the system in a majority of cases, as shown in Table~\ref{DigAttTable}. 

\smallskip
\noindent
\textbf{Choice of Dimensionality Reduction Techniques}. The effectiveness of dimensionality reduction techniques depends on the ability to learn group-level statistics of different types of paper sheets. 
Useful information may not be present along arbitrary directions, and projections on such directions can fail to learn effective digital forgeries. 
The successful use of PCA, commonly used to identify projection directions that capture the maximum variance in the data, suggests the potential to learn even more effective compression strategies, which could, in turn, help easily identify malicious samples in holdout datasets.

To explore better compression strategies, we implement manifold learning methods such as Uniform Manifold Approximation and Projection (UMAP) \cite{mcinnes2018umap}, which can better capture the lower-dimensional structure of the holdout set. We calculate the embeddings of our dataset using both PCA and UMAP, plotted in Fig.~\ref{fig:dimred}(c)--(d). We observe that UMAP embeddings exhibit distinct characteristics. Specifically, compared to PCA's principal components, the inter-class distance (i.e., the distance between different paper sheets) in UMAP embeddings is smaller, while the intra-class spread (i.e., the variation within samples of the same sheet) is larger. This suggests that UMAP may be even more effective at uncovering malicious samples in the holdout dataset, making the choice of dimensionality reduction technique significant. However, calculating the inverses of UMAP embeddings is computationally expensive and becomes problematic when applied to high-dimensional vectors, such as norm maps, limiting their direct use in digital forgery attacks. Future work should combine UMAP with neural compression techniques, such as autoencoders, to address these limitations.

\smallskip
\noindent
\textbf{Theoretical Likelihood of Collision at Initialization}. The best-case average complexity of a brute force attack is $10^{13}$ iterations (or the acquisition of $10^{13}$ holdout samples). Due to the exponential growth in search space with increasing dimensions—a phenomenon known as the ``curse of dimensionality''—the likelihood of a random 40k-d vector being sufficiently close to the reference norm map is extremely low. In fact, it is on the order of $10^{-20,916}$, effectively guaranteeing that the collision will almost never occur. We theoretically prove this argument in Appendix C of the supplemental material. In comparison, the best-case average complexity of efficient digital attacks (ranging from $10^1$--$10^2$) is 11--12 orders of magnitude lower than that of a brute force attack, highlighting their effectiveness.

\section{Toward Holistic Security: Strategy Space of Authentication Attacks\label{Sec5}}

Section~\ref{Sec4} provides the first experimental evidence of both physical and digital vulnerabilities in paper-PUF-based authentication systems, highlighting the gap between the system's theoretical design and practical deployability. In this section, we bridge this gap by systematically elaborating on potential system-level threats in paper-PUF-based authentication systems, guiding future research on secure system design. 
Leveraging our operational framework outlined in Section~\ref{Sec3} and illustrated in Fig.~\ref{fig:framework}, we conduct a stage-by-stage attack strategy analysis of the authentication process, delineating potential threats and revealing insights previously lacking from the literature. 
By integrating theories and techniques from biometrics, signal processing, and hardware PUF research, we identify key application-specific vulnerabilities, such as reverse engineering attacks, which have not been explored before. 
Unless otherwise stated, we focus on the verification phase.

\vspace{-2mm}

\subsection{Desired Security Properties}

Before service providers or stakeholders of the supply chain deploy a paper-PUF-based authentication system, it is essential to understand the key desired security properties. 
Table~\ref{fig:proptable} presents the desired security properties selected through a review of key literature sources, including the NIST specifications for biometric systems \cite{grother2013biometric} and seminal research on biometric authentication \cite{jain2006biometrics}. 
The table also indicates which properties are impacted when specific stages of the system are attacked.
We outline the desired security properties as follows:

\begin{enumerate}[left=0pt]
\itemsep=3pt 
\item \textbf{Confidentiality} \cite{nist_glossary}: Confidentiality deals with the covertness of the data in the system. For example, protecting the reference feature from leakage.
\item \textbf{Integrity} \cite{nist_glossary}: This refers to the resilience of data in the system to adversarial attempts so that authentication results can be trusted. Securing reference features in the database can help ensure this property.
\item \textbf{Revocability} \cite{nist_glossary}: Since intrinsic identifiers are the ``fingerprints'' of the products, it is critical to secure them by using cancelable and re-enrollable hashing methods.
\item \textbf{Replay Resistance} \cite{nist_glossary}: The verification system should be protected against adversaries who may try to replay old data and attack the channels between the different operational stages.
\item \textbf{Availability} \cite{nist_glossary}: Since even a small system outage can cause large losses for supply chain applications, the availability of the system is important. Being distributed and incorporating redundancy is crucial to mitigate DoS attacks. 
\end{enumerate}

\vspace{-2mm}

\begin{table}[!t]
    \caption{\label{fig:proptable} Properties targeted (\ding{52}) and not targeted (\rule[2pt]{4pt}{1.2pt}) in the stage-wise threat modeling of the operational framework}
    \resizebox{\linewidth}{!}{%
    \centering
    \renewcommand{\arraystretch}{1.2}
\begin{tabular}{P{1.5cm}*{5}{c}}
                                        \toprule
    &   \multicolumn{5}{c}{\thead{System Properties}}
                                    \\  \cmidrule{2-6}
\thead{Attacked \\ Stage}
    &   {\thead{Confiden-\\tiality}} 
        &   {\thead{Integrity}} 
            &   {\thead{Revo-\\cability}} 
                &   {\thead{Replay \\ Resistance}}
                    &   {\thead{Availability}}
                                    \\ \midrule
\centering 1
    & \rule[1pt]{4pt}{1.2pt} & \rule[1pt]{4pt}{1.2pt} & \rule[1pt]{4pt}{1.2pt} & \rule[1pt]{4pt}{1.2pt} & \ding{52}            \\
\centering 2
    & \ding{52}  & \ding{52}   & \rule[1pt]{4pt}{1.2pt} & \rule[1pt]{4pt}{1.2pt} & \rule[1pt]{4pt}{1.2pt}            \\
\centering 3
    & \ding{52}  & \ding{52}  & \ding{52}  & \ding{52} & \ding{52}             \\
\centering 4
     & \ding{52}  & \rule[1pt]{4pt}{1.2pt} & \rule[1pt]{4pt}{1.2pt} & \rule[1pt]{4pt}{1.2pt} & \rule[1pt]{4pt}{1.2pt}
                                    \\  \bottomrule
    \end{tabular}}
    \vspace{-4mm}
\end{table}

\subsection{\label{Attacks_Stage_1}Attacks on Stage 1: Image Acquisition and Preprocessing}

This stage employs an imaging sensor on a flatbed scanner or mobile camera to capture images of the paper surface, which are then preprocessed for feature extraction. 
We examine key threats to this stage, including physical denial-of-service (DoS) attacks that disrupt image acquisition and preprocessing, and sensor spoofing attacks that feed malicious inputs to authenticate counterfeit products.

\smallskip
\noindent \textbf{Security Properties Targeted}. \textit{Availability}

\smallskip
\noindent \textbf{Physical DoS Attacks}. 
Several works \cite{wong2017counterfeit, flatbed} use image alignment markers (left panel, top row of Fig.~\ref{fig:framework}) to align multiple snapshots of the same patch. An adversary can exploit this practice by tampering with the alignment markers, impairing authentication performance, or disrupting the entire authentication pipeline. Since subsequent stages rely on accurate image alignment, such tampering can render the authentication feature inaccurate or unusable. Additionally, paper PUFs reflect  the microstructures of the paper surface, making them susceptible to physical tampering attacks. As experimentally demonstrated in Sec.~\ref{Sec4.2}, an adversary may tear or scratch the paper surface, damaging the microstructures, or introduce significant alterations, such as smudging ink, printing patterns, or placing stickers over the PUF verification region, thereby sabotaging the system. Attacks like tearing or scratching are easy to execute and can effectively disrupt proper authentication completely.

\smallskip
\noindent \textbf{Physical Sensor Spoofing}.  
Spoofing involves injecting manipulated input samples into the acquisition module, generating artificially high decision scores that allow counterfeit products to pass authentication, akin to sensor spoofing in biometrics \cite{schuckers2002spoofing}. 
Traditional identifiers such as 1D/2D codes or RFID tags have been shown to be easily cloned \cite{cloneRFID, counterfeitcdp1}. 
This enables adversaries to clone identifiers associated with genuine products and associate the cloned codes/tags with the counterfeit product, spoofing traditional anti-counterfeiting solutions. 
The vulnerabilities of paper-PUF-based authentication systems to similar attacks remain underexplored. 
Motivated adversaries may adapt image-domain spoofing techniques from face-recognition systems \cite{kumar2017comparative, wu2020making} for paper-PUF authentication. These attacks could involve presenting manipulated images, such as printed photos of genuine products.
A potential easy spoofing attack against paper-PUFs could involve scanning genuine packaging and printing this scan onto counterfeit packaging. 
Authentication systems that rely strictly on extracting visual features from paper surfaces \cite{smith1999microstructure,beekhof2008secure} may be easily spoofed.
Such spoofed images could bypass the feature extraction module, compromising the authentication process.

\vspace{-2mm}
\subsection{\label{Attacks_Stage_2}Attacks on Stage 2: Feature Extraction Module}

The second stage is responsible for extracting the authentication features from the preprocessed images output by the first stage.
This stage is highly vulnerable to modeling attacks that bypass feature extraction, digital spoofing attacks that manipulate the feature extraction process, and synthetic feature generation attacks using generative models to create fake features that can pass authentication.

\smallskip
\noindent \textbf{Security Properties Targeted}. \textit{Confidentiality; Integrity}

\smallskip
\noindent \textbf{Surrogate Models}. Modeling surrogates has garnered significant attention in machine learning \cite{bhambri2019survey}, where the goal is to learn the underlying mathematical representation that maps the input space of a target system to its output space, using paired input--output data.
Machine learning-based modeling attacks have also been investigated in the context of arbiter PUFs \cite{pufattack2, ruhrmair2013puf}. Such attacks aim to learn the delay characteristics of arbiter PUFs from observed challenge--response pairs, assuming an underlying additive linear delay model. Prior studies have shown that even simple models, such as logistic regression, are capable of emulating the delay behavior of arbiter PUFs, as the problem can be viewed as learning the parameters of a separating hyperplane \cite{pufattack2}---a well-studied problem in machine learning. However, this representation may not correspond to the true underlying mechanism but could instead be an overfit mapping.
In contrast to arbiter PUFs, paper PUFs are high-dimensional, employing light as a challenge signal and using the captured images as observed responses. 
Nevertheless, estimating the stochastic microstructures through norm maps has also been shown to be equivalent to learning the parameters of a separating hyperplane \cite{wong2017counterfeit}, similar to arbiter PUFs. 
This similarity suggests that modeling attacks against paper PUFs may be feasible, and the ideas from arbiter PUFs could potentially transfer to this domain.

\smallskip
\noindent \textbf{Surrogate Models + Digital Spoofing}. In addition to training a surrogate feature extraction module, a motivated adversary may employ machine learning-based adversarial attacks on these surrogates to generate digital spoofs. For example, the adversary may collaborate with insiders to use optimization techniques such as projected gradient descent (PGD) \cite{madry2018towards} to ensure that the generated feature representation of a counterfeit product closely matches that of the genuine product, while the captured images look natural with minimal perturbations. Given the transferability of machine learning adversarial attacks \cite{papernot2016transferability}, attacks trained on surrogates are likely to be effective on original feature extractors.

\smallskip
\noindent \textbf{Synthetic Feature Generation}. 
Another machine-learning-based attack that can compromise the security of this stage is synthetic data generation. 
In the past decade, generative models have risen in popularity, learning the underlying latent space distribution of images \cite{goodfellow2014generative,ho2020denoising}. These distributions are sampled to generate semantically and visually realistic images. 
Since many authentication methods rely on visual or image-like physical features, generative models such as diffusion models \cite{ho2020denoising} could be used to generate fake features that pass the authentication process.
To achieve this, an adversary could modify the training scheme of generative models trained to generate synthetic digital paper PUF representations. 
By training on image--feature pairs and conditioning this process on additional information that allows the generated features to pass authentication, the adversary could potentially guide the model to create counterfeit features that circumvent authentication. 
This additional information may be obtained by repeatedly querying the decision-making system and using the observed similarity scores to fine-tune the generative model. 
Such strategies have demonstrated success in biometric systems, such as iris recognition \cite{kohli2017synthetic}, and may also be feasible on image-like paper PUFs.

\vspace{-2mm}
\subsection{\label{Attacks_Stage_3}Attacks on Stage 3: Storing Reference Data}  

The third stage involves storing template features in the reference database, which are later extracted and compared using the decision-making system. 
This stage is vulnerable to digital denial-of-service (DoS) attacks aiming to disable the reference database, template leakage attacks aiming to manipulate stored reference features, and reverse engineering attacks aiming to synthesize input images that estimate leaked templates.

\smallskip
\noindent \textbf{Security Properties Targeted}. \textit{Confidentiality; Integrity; Revocability; Replay Resistance; Availability;}

\smallskip
\noindent \textbf{Digital DoS Attacks}. An adversary can opt for a digital DoS attack affecting this stage such as resource overloading~\cite{adler2008biometric} by sending unnecessary queries through the first stage or by collaborating with insiders and directly querying the reference database.

\smallskip
\noindent \textbf{Template Leakage}. Adversaries may exploit sophisticated attacks or insider knowledge, such as collusion or social engineering, to compromise the reference database. If they gain read-and-write access, they could insert a crafted feature corresponding to a counterfeit product, associate it with the correct product ID, and write it to the database, allowing the counterfeit product to be authenticated as genuine. Similar attacks can occur during the Registration phase (discussed in Sec.~\ref{Sec3}), acting as a backdoor to facilitate continuous misauthentication. If only read access is obtained, adversaries may launch a data extraction attack, retrieve the genuine product's template feature, and replay it to the decision-making system, bypassing authentication.

\smallskip
\noindent \textbf{Template Leakage + Reverse Engineering}. If the adversary cannot directly inject the template into the decision-making system (i.e., if the integrity of the channels is preserved) and can only interact with the first stage, they may need to synthesize input images that recreate the leaked template. This creates a reverse engineering problem. If the adversary successfully converts the leaked template into plausible input images, they can feed these into the authentication system to bypass authentication. While PUF responses are generally considered non-invertible and unclonable due to the inherent randomness in their generation \cite{clarkson2009fingerprinting, wongicip, wong2017counterfeit, flatbed}, this non-invertibility may be compromised since the adversary can access the template features. Calculating pixel intensity values from leaked PUF templates is feasible using the fully diffuse reflection model \cite{wong2017counterfeit} by modeling environmental factors, such as the direction and intensity of incident light. 

\vspace{-1mm}
\subsection{\label{Attacks_Stage_4}Attacks on Stage 4: Decision-Making System}  

The final stage is the decision-making system, which compares the test feature with the reference templates stored in the database to make the final authentication decision. This stage is vulnerable to hill-climbing attacks, iteratively guessing a test feature to pass authentication. 

\smallskip
\noindent \textbf{Security Properties Targeted}. \textit{Confidentiality}

\smallskip
\noindent \textbf{Hill Climbing Attacks}. The choice of the distance metric used to evaluate a match also has security ramifications. There is a large body of work in the field of biometrics that has investigated ``information leaking metrics" 
\cite{rathgeb2010attacking}. 
If such metrics are used in the decision-making stage~$\delta$ and the distance scores are observable, some information is leaked and hill-climbing attacks \cite{galbally2007bayesian} are feasible. 
While manufacturing another paper sheet with identical PUFs may seem infeasible due to the complex intertwining of cellulose fibers, circumventing authentication by learning digital forgeries of PUF features (e.g., norm maps) can be highly effective, as demonstrated in Sec.~\ref {Sec4.3}.

\section{Conclusion \label{Sec8}}

As global supply chains undergo restructuring and the demand for physical goods continues to rise, there is an urgent need for robust solutions to combat counterfeiting and protect public health and safety. 
In this paper, we investigate the security of paper-PUF-based authentication systems, an economical and high-performance anti-counterfeiting solution, by exposing vulnerabilities in both the physical and digital domains. 
Despite the inherent unclonability of paper PUFs, we have demonstrated that without proactive safeguards, these systems are susceptible to physical denial-of-service and digital forgery attacks. 
Adversaries can either physically tamper with the paper packaging to sabotage authentication or generate digital forgeries of paper PUFs to circumvent authentication.
Our holistic system-level analysis further identifies key attack strategies and proposes countermeasures to enhance security. 
Future anti-counterfeiting solutions should build upon our findings, incorporating stronger defenses such as improved correlators, revocable hashing, and homomorphic encryption to mitigate discovered vulnerabilities.

\bibliographystyle{IEEEtran}

\bibliography{IEEEabrv,references}

\balance

\newpage

\section*{Supplemental Document}

\appendices

\section{\label{App:A_DoS}Extended Results for Physical DoS Attacks}

\begin{table}[!htb]
\centering
\caption{Effect of Physical DoS Attacks at 5\%, 10\%, 25\%, 50\%, and 75\% Strengths on Matched Case Correlation.}
{\renewcommand{\arraystretch}{1.15}
\begin{tabular}{cccc}
\toprule
\textbf{\begin{tabular}[c]{@{}c@{}}Physical \\ Attack Setup\end{tabular}} & \textbf{\begin{tabular}[c]{@{}c@{}}Attack \\ Strength (\%)\end{tabular}} & \textbf{\begin{tabular}[c]{@{}c@{}}$\boldsymbol{x}$ Corr Coef\\ Mean $\downarrow$ (Std)\end{tabular}} & \textbf{\begin{tabular}[c]{@{}c@{}}$\boldsymbol{y}$ Corr Coef\\ Mean $\downarrow$ (Std)\end{tabular}} \\ \hline
\textbf{No Attack} & N/A & 0.46 (0.08) & 0.36 (0.09) \\ \hline 
\multirow{5}{*}{\textbf{Scratching}} & 5 & 0.37 (0.02) & 0.16 (0.07) \\ \cline{2-4} 
 & 10 & 0.28 (0.03) & 0.10 (0.05) \\ \cline{2-4} 
 & 25 & 0.16 (0.02) & 0.05 (0.04) \\ \cline{2-4} 
 & 50 & 0.07 (0.02) & 0.03 (0.02) \\ \cline{2-4} 
 & 75 & 0.03 (0.01) & 0.01 (0.01) \\ \hline
\multirow{5}{*}{\textbf{\begin{tabular}[c]{@{}c@{}}Physical \\ Patch Attack\end{tabular}}} & 5 & 0.45 (0.03) & 0.17 (0.04) \\ \cline{2-4} 
 & 10 & 0.34 (0.04) & 0.13 (0.06) \\ \cline{2-4} 
 & 25 & 0.20 (0.02) & 0.09 (0.03) \\ \cline{2-4} 
 & 50 & 0.09 (0.01) & 0.04 (0.02) \\ \cline{2-4} 
 & 75 & 0.03 (0.00) & 0.01 (0.01) \\ \hline
\multirow{5}{*}{\textbf{Scribbling}} & 5 & 0.20 (0.02) & 0.13 (0.03) \\ \cline{2-4} 
 & 10 & 0.14 (0.02) & 0.06 (0.01) \\ \cline{2-4} 
 & 25 & 0.09 (0.02) & 0.02 (0.01) \\ \cline{2-4} 
 & 50 & 0.03 (0.01) & 0.01 (0.01) \\ \cline{2-4} 
 & 75 & 0.01 (0.01) & 0.01 (0.01) \\ \hline
\multirow{2}{*}{\textbf{Crumpling}} & \begin{tabular}[c]{@{}c@{}}Random\end{tabular} & 0.03 (0.02) & 0.06 (0.03) \\ \cline{2-4} 
 & \multicolumn{1}{c}{Folding} & 0.02 (0.01) & 0.08 (0.02) \\ 
\bottomrule
\multicolumn{4}{l}{``$\downarrow$'' indicates that a lower mean is desired by attackers aiming to} \\
\multicolumn{4}{l}{sabotage authentication.}
\end{tabular}}
\vspace{-4mm}
\end{table}

\section{\label{appB_digital}Algorithms for Digital Forgery Attacks}

\begin{algorithm}
\SetAlgoLined
\KwIn{Initial norm map guess $\n_0$, noise range $\delta$, similarity threshold $\tau$, $T$ iterations}
\KwOut{Learned synthetic norm map $\n_T$}

\textbf{Initialize:} Set $t = 0$, and query similarity score $\rho_0$ between $\n_0$ and the reference norm map\;

\While{$\rho_t < \tau$ and $t < T$}{
    Randomly select a subset of dimensions from $\n_{t-1}$\;
    
    Perturb selected dimensions by adding noise uniformly sampled from $[-\delta, \delta]$ to form $\n_t$\;
    
    Query similarity score $\rho_t$ between $\n_t$ and the reference norm map\;
    
    \If{$\rho_t \geq \rho_{t-1}$}{
        Accept $\n_t$ as the current best guess\;
    }
    \Else{
        Set $\n_t = \n_{t-1}$ (retain previous guess)\;
    }
    
    Increment iteration count $t \gets t + 1$\;
}
\textbf{Return} $\n_T$
\caption{Baseline Greedy Attack}
\end{algorithm}

\begin{algorithm}
\SetAlgoLined
\KwIn{Initial norm map guess $\n_0$, noise range $\delta$, similarity threshold $\tau$, $T$ iterations, hold-out dataset $\mathcal{D}_{\text{hold}}$}
\KwOut{Learned synthetic norm map $\n_T$}

\textbf{Step 1: Dimensionality Reduction }\;
Learn a transformation $h$ on $\mathcal{D}_{\text{hold}}$ that compresses the norm maps into a low-dimensional latent space\;
Apply the transformation to $\n_0$, resulting in a compressed latent vector $\mathbf{z}_0 = h(\n_0)$\;

\textbf{Step 2: Initialization}\;
Set $t = 0$, and query the similarity score $\rho_0$ between $\n_0$ and the reference norm map\;

\While{$\rho_t < \tau$ and $t < T$}{
    Randomly select a subset of dimensions from the compressed vector $\mathbf{z}_{t-1}$\;
    
    Perturb selected dimensions by adding noise uniformly sampled from $[-\delta, \delta]$ in the latent space to form $\mathbf{z}_t$\;
    
    Use transformation $h^{-1}$ to map $\mathbf{z}_t$ back to the original feature space, obtaining $\n_t = h^{-1}(\mathbf{z}_t)$\;
    
    Query the similarity score $\rho_t$ between $\n_t$ and the reference norm map\;
    
    \If{$\rho_t \geq \rho_{t-1}$}{
        Accept $\mathbf{z}_t$ as the current best guess in the latent space\;
    }
    \Else{
        Set $\mathbf{z}_t = \mathbf{z}_{t-1}$ (retain previous guess)\;
    }
    
    Increment iteration count $t \gets t + 1$\;
}

\textbf{Step 3: Black-box Optimization}\;
For faster convergence, apply black-box optimization techniques (e.g., Nelder--Mead, Powell, or conjugate-gradient) in latent space $\mathbf{z}$ directly\;

\textbf{Return} $\n_T = h^{-1}(\mathbf{z}_T)$
\caption{Efficient Digital Attacks}
\end{algorithm}

\nobalance

\section{\label{App:C}Likelihood of collision at initialization}

\noindent
\textbf{Proposition:}  
Let \( \mathbf{M}_{\text{rand}} \in \mathbb{R}^d \) be a random uniformly sampled norm map instantiation, and \( \mathbf{M}_{\text{ref}} \in \mathbb{R}^d \) be the reference norm map. 
Let $\max \|\mathbf{M}_{\text{rand}}\| = R$. 
The probability \( p \) of initializing a \( \mathbf{M}_{\text{rand}} \) that satisfies \( \|\mathbf{M}_{\text{rand}} - \mathbf{M}_{\text{ref}}\| \leq \epsilon \), where \( \epsilon > 0 \) is
\[
p = \left( \frac{\epsilon}{R} \right)^d.
\] 
\noindent
\begin{proof}
We need to find the probability of the event
\begin{equation}
    \label{eq:1}
     \|\mathbf{M}_{\text{rand}} - \mathbf{M}_{\text{ref}}\| \leq \epsilon.
\end{equation} 
This event implies that $\mathbf{M}_{\text{rand}}$ is within an $\epsilon$ radius of $\mathbf{M}_{\text{ref}}$ in \( \mathbb{R}^d \). The volume of a hypersphere of radius $r$ in \( \mathbb{R}^d \) is given by
\begin{equation}
    V(r) = \frac{\pi^{d/2} r^d}{\Gamma(d/2 + 1)}.
\end{equation}
where $\Gamma(\cdot)$ is the Gamma function. Substituting \( r = \epsilon \), the volume of the region that satisfies \eqref{eq:1} is
\begin{equation}
    V_{\text{event}} = \frac{\pi^{d/2} \epsilon^d}{\Gamma(d/2 + 1)}.
\end{equation} 
The random vector \( \mathbf{M}_{\text{rand}} \) is sampled uniformly from a larger hypersphere of radius \( R \) centered at the origin. The volume of this hypersphere \( V_{\text{event}} \) is
\begin{equation}
    V_{\text{total}} = \frac{\pi^{d/2} R^d}{\Gamma(d/2 + 1)}.
\end{equation}
Finally, since \( \mathbf{M}_{\text{rand}} \) is sampled uniformly, the probability of it lying within $V_{\text{event}}$ is proportional to the ratio of the volumes
\begin{align}
     p &= \frac{V_{\text{event}}}{V_{\text{total}}} = \left( \frac{\epsilon}{R} \right)^d. 
\end{align} 
\end{proof}
\noindent
\textbf{Probability of Collision:} 
The probability of a 40,000-dimensional random norm map ($d$ = 40,000) being sufficiently close ($\epsilon = 0.3$) to the reference norm map, assuming each element in the random vector is normalized in the range [0,1] ($R$ = 1), is:
\begin{equation}
    p = \left( \frac{0.3}{1} \right)^{40,000} \approx 7.02 \times 10^{-20916}.
\end{equation}

\end{document}